\documentclass[10pt,twocolumn,letterpaper]{article}

\usepackage[pagenumbers]{cvpr} 
\usepackage{graphicx}
\usepackage{amsmath}
\usepackage{amssymb}
\usepackage{booktabs}

\usepackage{subfiles}
\usepackage{subcaption}
\usepackage{xcolor}
\usepackage{arydshln}
\usepackage[export]{adjustbox}

\makeatletter
\@namedef{ver@everyshi.sty}{}
\makeatother
\usepackage{tikz}
\usetikzlibrary{positioning}

\def\wacvPaperID{260} 

\usepackage[pagebackref,breaklinks,colorlinks]{hyperref}

\def\LRA{\Leftrightarrow}
\def\EqZx#1{ \mathbb{E}_{q_{Z|x}}\left[ #1 \right] }
\def\qzi{ q_i }
\def\pzi{ p_i }
\def\ourblock{latent block}
\def\ours{QRes-VAE}
\def\oursl{\ours{} (34M)}
\def\ourss{\ours{} (17M)}
\def\ourslossless{\ours{} (lossless)}

\begin{document}

\title{Lossy Image Compression with Quantized Hierarchical VAEs}

\author{
Zhihao Duan\textsuperscript{*} \quad \quad
Ming Lu\textsuperscript{\textdagger} \quad \quad
Zhan Ma\textsuperscript{\textdagger} \quad \quad
Fengqing Zhu\textsuperscript{*}
\\
\textsuperscript{*} Purdue University, West Lafayette, Indiana, U.S. \\
\textsuperscript{\textdagger} Nanjing University, Nanjing, Jiangsu, China \\
{\tt\small duan90@purdue.edu, luming@smail.nju.edu.cn, mazhan@nju.edu.cn, zhu0@purdue.edu}
\thanks{Dr. Z. Ma is partially funded by Microsoft Research Asia.}
}

\maketitle
\thispagestyle{empty}

\begin{abstract}
Recent research has shown a strong theoretical connection between variational autoencoders (VAEs) and the rate-distortion theory. Motivated by this, we consider the problem of lossy image compression from the perspective of generative modeling.
Starting with ResNet VAEs, which are originally designed for data (image) distribution modeling, we redesign their latent variable model using a quantization-aware posterior and prior, enabling easy quantization and entropy coding at test time.
Along with improved neural network architecture, we present a powerful and efficient model that outperforms previous methods on natural image lossy compression.
Our model compresses images in a coarse-to-fine fashion and supports parallel encoding and decoding, leading to fast execution on GPUs.
Code is made \href{https://github.com/duanzhiihao/lossy-vae}{available online}.
\end{abstract}


\section{Introduction}


Data (in our context, image) compression and generative modeling are two fundamentally related tasks.
Intuitively, the essence of compression is to find all ``patterns" in the data and assign fewer bits to more frequent patterns.
To know exactly how frequent each pattern occurs, one would need a good probabilistic model of the data distribution, which coincides with the objective of (likelihood-based) generative modeling.
This connection between compression and generative modeling has been well established, both theoretically and experimentally, for the \textit{lossless} setting. In fact, many modern image generative models are also best-performing lossless image compressors~\cite{townsend2020hilloc, zhang2021nelloc}.


A similar connection can be drawn for the \textit{lossy} compression setting.
In particular, a popular class of image generative models, variational autoencoders (VAEs)~\cite{kingma14vae}, has been proved to have a rate distortion (R-D) theory interpretation~\cite{alemi2018fixing_elbo, yang2022sandwich}.
With a distortion metric specified, VAEs learn to ``compress" data by minimizing a \textit{tight} upper bound on their information R-D function~\cite{yang2022sandwich}, showing great potential for application to lossy image compression.
However, existing best-performing VAEs~\cite{vahdat2020nvae, child2021vdvae} employ continuous latent variables, which cannot be straightforwardly coded into bits, and thus cannot be used for practical image compression.
Although several methods have been developed to turn VAEs into practical codecs, \eg, by communicating samples~\cite{flamich2020rec} and post-training quantization~\cite{yang2020quantization}, neither achieved satisfactory R-D performance compared to existing methods such as VVC intra codec~\cite{pfaff2021vvc_intra}.


Despite the lack of a practical coding algorithm, the
potential of VAEs in lossy compression
has also been reflected in the image coding community. Although independently developed from the perspective of \textit{transform coding}, many learning-based image compressors resemble a simple VAE in which the latent variables are first-order Markov~\cite{balle18hyperprior}.
Given that such simple VAEs are shown to be suboptimal in generative image modeling~\cite{Sonderby2016laddervae}, we hypothesize that a more powerful VAE architecture, \eg, \textit{hierarchical VAEs}, would also achieve a better lossy compression performance.
\begin{figure}[t]
    \includegraphics[width=0.98\linewidth]{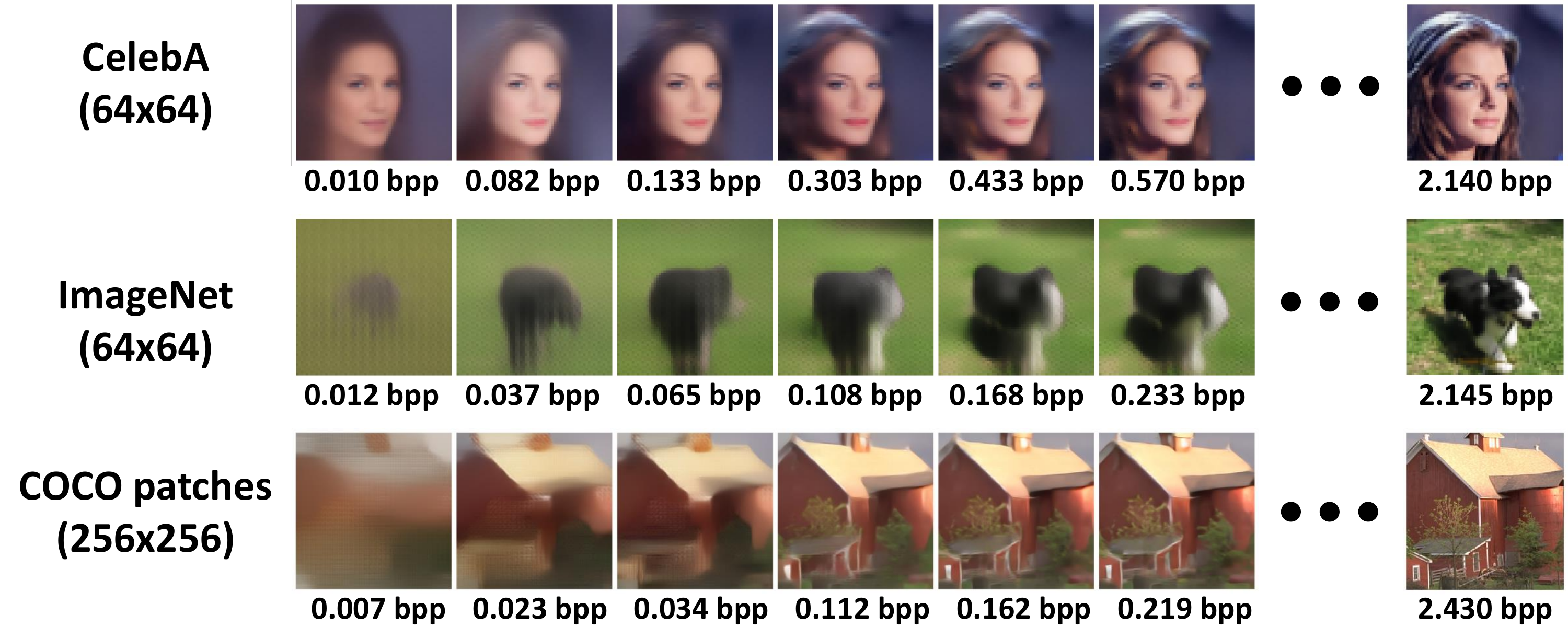}
    \caption{
    Our \textbf{\ours{}} (for Quantized ResNet VAE) model learns a deep hierarchy of features and compresses/decompresses images in a coarse-to-fine fashion. Models are separately trained on each dataset.
    }
    \label{fig:intro_progressive}
\end{figure}

Motivated by this, we adopt hierarchical VAE architectures that are originally designed for generative image modeling and use them for lossy compression.
We redesign their probabilistic model to allow easy quantization and practical entropy coding, in a way similar to popular lossy compression methods~\cite{balle18hyperprior, minnen2018joint}.
Specifically, we start from a powerful hierarchical VAE architecture, \textit{ResNet VAEs}~\cite{kingma2016iafvae}, and introduce modifications including 1) a uniform posterior, 2) a Gaussian convolved with uniform prior, and 3) revised network architectures.
Our new model, \textbf{\ours{}} (for quantized ResNet VAE), achieves better R-D performance on natural image compression than existing lossy compression methods.
Furthermore, our model compresses images in a coarse-to-fine manner (Fig.~\ref{fig:intro_progressive}) thanks to its hierarchical architecture, while avoids the slow sequential encoding/decoding as experienced by spatially autoregressive image models~\cite{minnen2018joint}.


Our contributions are summarized as follows.
We propose to use a quantization-aware latent variable model for modern hierarchical VAEs, making practical entropy coding feasible.
We present a powerful and efficient VAE model that outperforms previous hand-crafted and learning-based methods on lossy image compression.
Our method narrows the gap between image compression and generation, at the same time providing insights into designing better image compression systems.
\section{Background and Related Work}
\label{sec:related}
In this section, we briefly summarize the preliminaries, introduce our notation, and review previous works.

\subsection{Variational Autoencoder (VAE)}
\label{sec:related_vae}
\begin{figure}[t]
\centering
    \begin{subfigure}[b]{0.23\textwidth}
        \centering
            \begin{tikzpicture}[
            rv_node/.style={circle, draw=black!100, thick, minimum size=8mm},
            ob_node/.style={circle, draw=black!100, thick, minimum size=8mm, fill=black!24},
            ]
            \node[ob_node] (x) {$X$};
            \node[rv_node] (z) [above=of x, label=right:$\sim p_Z(\cdot)$] {$Z$};
            \draw [-latex, blue, dashed] (x.north west) to [out=135,in=225] node[left=0mm] {$q_{Z|X}(\cdot)$} (z.south west);
            \draw[-latex] (z.south) -- (x.north) node [midway, right] (emission) {$p_{X|Z}(\cdot)$};
            
            \matrix [below left] at (2,4) {
              \draw[-latex, blue, dashed] (0,-0.24) -- (0.64,-0.24); & \node{\small Inference (encoding)};\\
              \draw[-latex] (0,-0.24) -- (0.64,-0.24); & \node{\small Sampling (decoding)};\\
            };
            \end{tikzpicture}
        \caption{VAE}
        \label{fig:vae_model_one}
    \end{subfigure}
    \hfill
    \begin{subfigure}[b]{0.23\textwidth}
        \centering
            \begin{tikzpicture}[
            rv_node/.style={circle, draw=black!100, thick, minimum size=8mm},
            ob_node/.style={circle, draw=black!100, thick, minimum size=8mm, fill=black!24},
            ]
            \node[ob_node] (x) {$X$};
            \node[rv_node] (z2) [above=of x] {$Z_2$};
            \node[rv_node] (z1) [above=of z2] {$Z_1$};
            \draw [-latex, blue, dashed] (x.north west) to [out=135,in=225] (z2.south west);
            \draw [-latex, blue, dashed] (x.north west) to [out=150,in=210] (z1.south west);
            \draw [-latex, blue, dashed] (z1.south west) to [out=225,in=135] (z2.north west);
            \draw[-latex] (z1.south) -- (z2.north);
            \draw[-latex] (z2.south) -- (x.north);
            \draw [-latex] (z1.south east) to [out=-45,in=45] (x.north east);
            \end{tikzpicture}
        \caption{2-layer ResNet VAE~\cite{kingma2016iafvae}}
        \label{fig:vae_model_res}
    \end{subfigure}
    \vspace{-0.2cm}
\caption{\textbf{Probabilistic model of VAEs}, where $X$ represents data, and $Z$ denotes latent variable(s). In this work, we use a 12-layer ResNet VAE for image compression.
}
\label{fig:vae_model}
\end{figure}
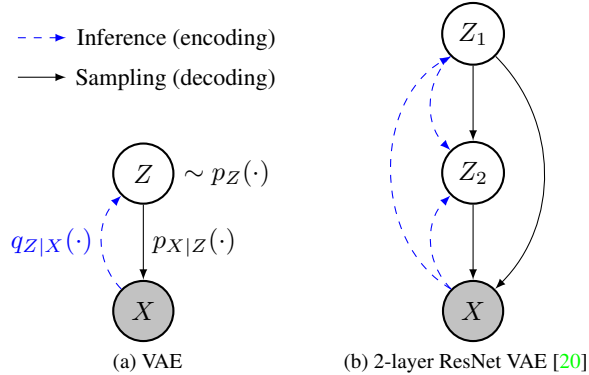

Let $X$ be the random variable representing data (in our case, natural images), with an unknown data distribution $p_\text{data}(\cdot)$.
VAEs~\cite{kingma14vae} model the data distribution by assuming a latent variable model as shown in Fig.~\ref{fig:vae_model_one}, {where $Z$ is the assumed latent variable.}
{In VAEs,} there is a prior $p_Z(\cdot)$, a {conditional data likelihood} (or decoder) $p_{X|Z}(\cdot)$ for sampling, and an approximate posterior (or encoder) $q_{Z|X}(\cdot)$ for variational inference.
Note that we omit the model parameters for simplified notation.

The training objective of VAE is to minimize a variational upper bound on the negative log-likelihood:
\begin{equation}
\label{eq:vae_loss}
\begin{aligned}
\mathcal{L}
&=
D_{\text{KL}}(q_{Z|x} \parallel p_Z) + \EqZx{ \log \frac{1}{p_{X|Z}(x | Z)} }
\\
&\ge -\log p_X(x),
\end{aligned}
\end{equation}
where $x$ is a an image, $q_{Z|x}$ is a shorthand notation for $q_{Z|X}(\cdot \, | x)$, and the minimization of $\mathcal{L}$ is w.r.t. the model parameters of $q_{Z|X}(\cdot)$, $p_{Z}(\cdot)$, and $p_{X|Z}(\cdot)$.


\textbf{Hierarchical VAEs:} {To improve the flexibility of VAEs}, the latent variable is often divided into disjoint groups, $Z \triangleq \{Z_1, Z_2, ..., Z_N\}$ {where $N$ is the number of groups}, to form a \textit{hierarchical VAE}.
Many best-performing VAEs~\cite{vahdat2020nvae, child2021vdvae, sinha2021crvae} follow a ResNet VAE~\cite{kingma2016iafvae} architecture, whose probabilistic model is shown in Fig.~\ref{fig:vae_model_res}.
During sampling, each latent variable $Z_i$ is conditionally dependent on all previous variables $Z_{<i}$, where $i \in \{1, 2, ..., N\}$ is the index, and during inference, $Z_i$ is also conditionally dependent on $X$ in addition to $Z_{<i}$.
For notational simplicity, we define
\begin{equation}
\label{eq:notation}
\begin{aligned}
\qzi (\cdot) & \triangleq q_{Z_i|X,Z_{<i}} (\cdot)
\\
\pzi (\cdot) & \triangleq p_{Z_i|Z_{<i}} (\cdot)
\end{aligned}
\end{equation}
to denote the approximate posterior and prior distributions for $Z_i$, respectively.
We also define $Z_{<1}$ to be an empty set, so we have $p_{1}(\cdot) \triangleq p_{Z_1}(\cdot)$ and $q_{1}(\cdot) \triangleq q_{Z_1|X}(\cdot)$.

\textbf{Discrete VAEs}:
A number of works have been proposed to develop VAEs with a discrete latent space. Among them, vector quantized VAEs (VQ-VAE)~\cite{oord2017vqvae, razavi2019vqvae2} are able to generate high-fidelity images, but they rely on a separately learned prior and cannot be trained end-to-end. Several works extend VQ-VAEs to form a hierarchy~\cite{williams2020hqa, willetts2020relaxed}, but they suffer from the same problem as VQ-VAE.
Another line of work, Discrete VAEs (DVAE)~\cite{rolfe2016dvae, vahdat2018dvaepp, vahdat2018dvaehash}, assumes binary latent variables and use Boltzmann machines as the prior.
However, none of them are able to scale to high resolution images.
In this work, we tackle these issues through {test-time} quantization.

\textbf{Data compression with VAEs:}
VAEs have a very strong theoretical connection to data compression.
When $X$ is discrete, VAEs can be directly used for lossless compression~\cite{townsend2018bbans, kingma2019bitswap} using the bits-back coding algorithm~\cite{hinton1993bitsback}.
When $X$ is continuously valued, the VAE objective (Eq.~\ref{eq:vae_loss}) has a rate-distortion (R-D) theory interpretation~\cite{alemi2018fixing_elbo} and has been used to estimate the information R-D function for natural images~\cite{yang2022sandwich}.
However, there lacks an entropy coding algorithm to turn VAEs into practical lossy coders.
Various works~\cite{agustsson2020universally, flamich2020rec, theis2022algorithms_comm_sapmles, flamich2022a_star_rec} have been conducted towards this goal, but they either require an intractable execution time or do not achieve competitive performance compared to existing lossy image coders.
In this work, we provide another pathway for turning VAEs into practical coders, \ie, by using a quantization-aware probabilistic model for latent variables.

\subsection{Lossy Image Compression}
\label{sec:related_lic}
Most existing lossy image coders follows the \textit{transform coding}~\cite{goyal2001tc} paradigm. 
Conventional, handcrafted image codecs such as JPEG~\cite{wallace1992jpeg}, JPEG 2000~\cite{Skodras2001jpeg2000}, BPG~\cite{lainema2012hevc_intra}, VVC intra~\cite{pfaff2021vvc_intra} use orthogonal linear transformations such as discrete cosine transform (DCT) and discrete wavelet transform (DWT) to decorrelate the image pixels before quantization and coding.
Learning-based coders achieve this in a non-linear way~\cite{balle2021nonlinear} by implementing the transformations and an entropy model using neural networks.
To improve compression efficiency, most previous works either explore efficient neural network blocks, such as residual networks~\cite{cheng2020cvpr}, self-attention~\cite{chen2021nlaic}, and transformers~\cite{lu2022tic, zou2022stf}, or develop more expressive entropy models, such as hierarchical~\cite{balle18hyperprior} and autoregressive models~\cite{minnen2018joint, minnen2020channelwise, he2021checkerboard}.

{
Interestingly, the learned transform coding paradigm can be equivalently viewed as a simple VAE~\cite{theis2017lossy, balle18hyperprior}, where the encoder (with additive noise), entropy model, and decoder in transform coding correspond to the posterior, prior, and conditional data likelihood in VAEs, respectively.
Along this direction, we propose to use a more powerful hierarchical VAE architecture for lossy image compression.
}

\section{Quantization-Aware Hierarchical VAE}
\label{sec:method}

\begin{figure*}[ht]
    \begin{subfigure}[b]{0.24\textwidth}
        \centering
        \includegraphics[width=\linewidth]{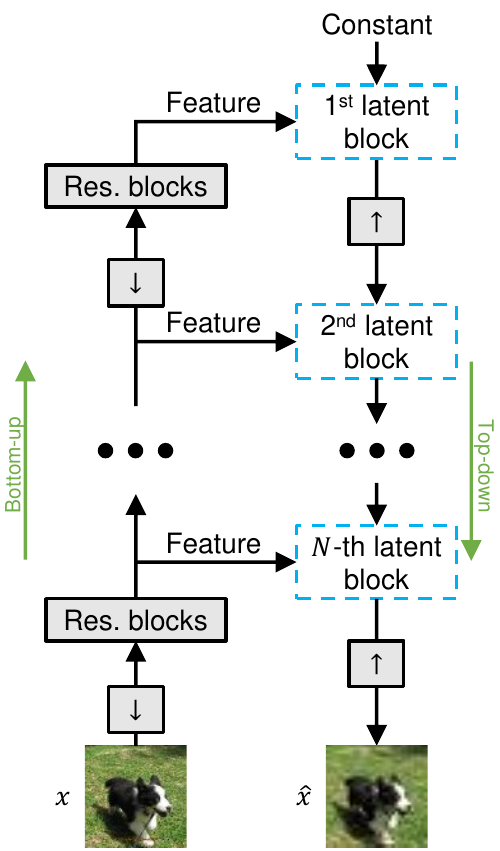}
        \caption{General framework}
        \label{fig:main_overall}
    \end{subfigure}
    \hfill
    \begin{subfigure}[b]{0.24\textwidth}
        \centering
        \includegraphics[width=\linewidth]{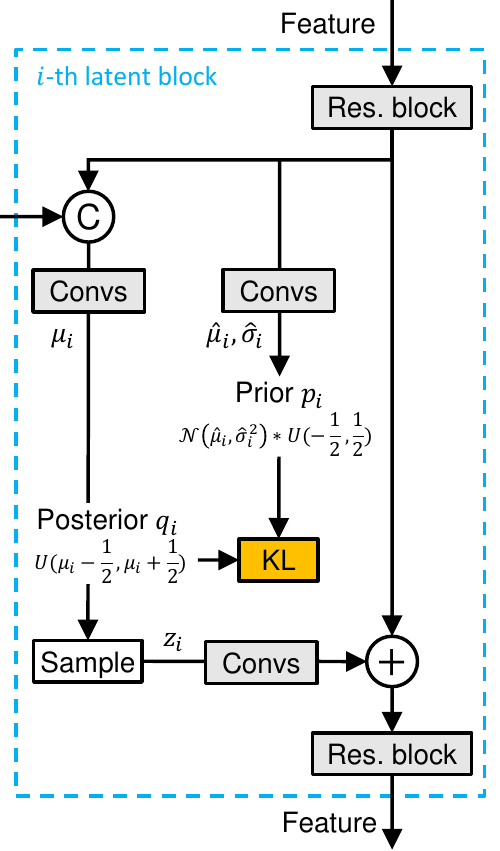}
        \caption{Latent block (training)}
        \label{fig:main_train}
    \end{subfigure}
    \hfill
    \begin{subfigure}[b]{0.24\textwidth}
        \centering
        \includegraphics[width=\linewidth]{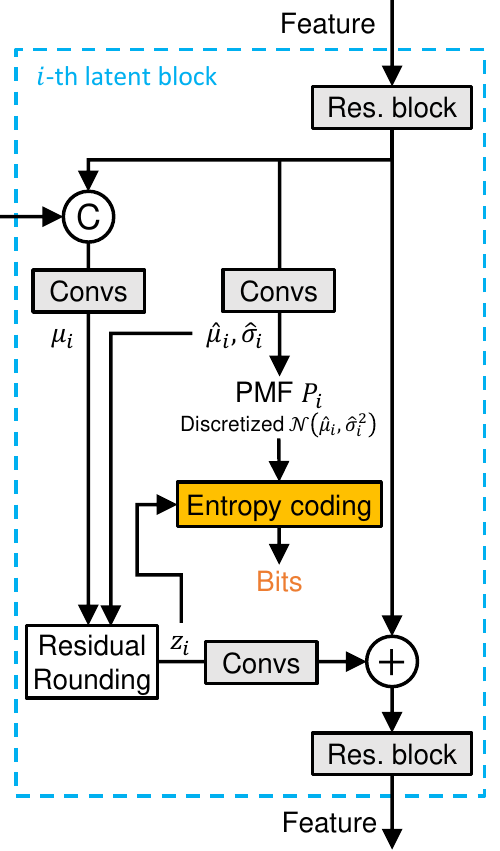}
        \caption{Latent block (compress)}
        \label{fig:main_encode}
    \end{subfigure}
    \hfill
    \begin{subfigure}[b]{0.24\textwidth}
        \centering
        \includegraphics[width=\linewidth]{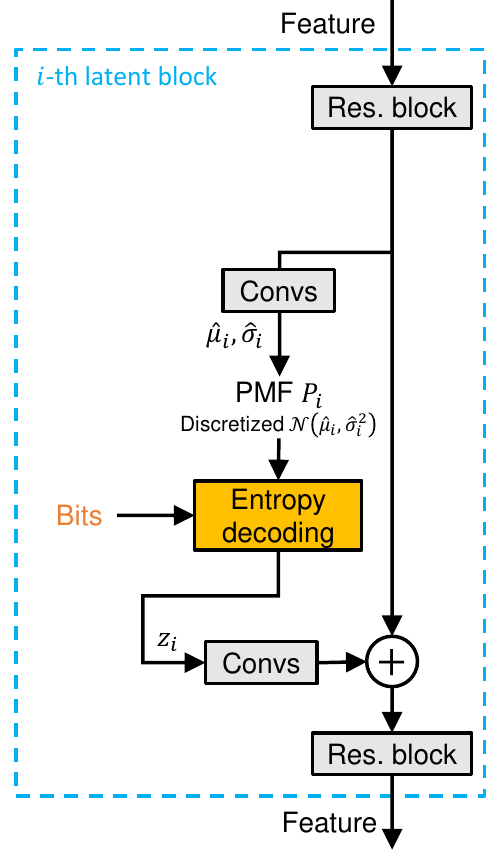}
        \caption{Latent block (decompress)}
        \label{fig:main_decode}
    \end{subfigure}
    \hfill
    \vspace{-0.2cm}
    \caption{\textbf{Overview of our quantized ResNet VAE (\ours{})}.
    In (a), we show the general framework of ResNet VAEs~\cite{kingma2016iafvae}. In our implementation, we use a 12-layer (\ie, 12 \ourblock{}s) architecture for \ours{}. The gray blocks represent neural networks, $\uparrow$ and $\downarrow$ denote upsampling and downsampling operations, respectively, \textcircled{c} denotes feature concatenation, and \textcircled{+} denotes addition.
    {The \ourblock{}s behave differently for training, compression, and decompression, which are illustrated in (b), (c), and (d), respectively. See Sec.~\ref{sec:method} for detailed description.}
    }
    \label{fig:main}
\end{figure*}



In this section, we first present our overall network architecture in Sec.~\ref{sec:method_architecture}.
{Next, we describe our probabilistic model and loss function in Sec.~\ref{sec:method_model}.}
We detail how to implement practical image coding in Sec.~\ref{sec:method_compression}.
Finally, we provide discussion in Sec.~\ref{sec:method_discussion}, highlighting the {relationship of our approach to previous methods.}

\subsection{Network Architecture}
\label{sec:method_architecture}
Our overall architecture is similar to the framework of ResNet VAEs~\cite{kingma2016iafvae, vahdat2020nvae, child2021vdvae}, on which we made modifications to adapt them for practical compression.
We briefly describe our network architecture in this section and refer readers to the Appendix~\ref{sec:appendix_detail_architecture} and our code for full implementation details.

\textbf{Overall architecture:} Fig.~\ref{fig:main_overall} overviews the general framework of ResNet VAE, which consists of a bottom-up path and a top-down path.
Given an input image $x$, the bottom-up path produces a set of deterministic features, which are subsequently sent to the top-down path for (approximate) inference.
In the top-down path, the model starts with a learnable constant feature and then pass it into a sequence of \textit{\ourblock{}s},
where each \ourblock{} adds ``information", carried by the latent variables $z_i$, into the feature.
At the end of the top-down path, a reconstruction $\hat{x}$ is predicted by an upsampling layer.
Our \ourblock{}s behaves differently for training, compression, and decompression, which we describe in details in Sec.~\ref{sec:method_model} and Sec.~\ref{sec:method_compression}.

\textbf{Network components:}
We use patch embedding~\cite{dosovitskiy2021vit} (\ie, convolutional layer with the stride equal to kernel size) for downsampling operations.
We use sub-pixel convolution~\cite{shi2016subpixel} (\ie, 1x1 convolution followed by pixel shuffling) for upsampling, which can be viewed as an ``inverse" of the patch embedding layer.
Our choices for the downsampling/upsampling layers are motivated by their success in computer vision tasks~\cite{dosovitskiy2021vit, shi2016subpixel} as well as their efficiency compared to overlapped convolutions.
The residual blocks in our network can be chosen arbitrarily in a plug-and-play fashion.
We use the ConvNeXt~\cite{liu2022convnext} block as our choice, since we empirically found it achieves better performance than alternatives (see ablation study in Sec.~\ref{sec:exp_ablation}).


\subsection{Probabilistic Model and Training Objective}
\label{sec:method_model}
As we mentioned in Sec.~\ref{sec:related}, VAEs with continuous latent variables cannot be straightforwardly used for lossy compression (due to the lack of an entropy coding algorithm).
We also note that a combination of a) uniform quantization at test time and b) additive uniform noise at training time allows easy entropy coding, as well as having an elegant VAE interpretation~\cite{balle2016end2end, balle18hyperprior}.
{Inspired by this, we redesign the probabilistic model in ResNet VAEs, using uniform posteriors to enable quantization-aware training. For the prior, we use a Gaussian distribution convolved with uniform distribution, which has enough flexibility to match the posterior~\cite{minnen2018joint}.}
Note that under this configuration, the probabilistic model in each \ourblock{} closely resembles the mean \& scale hyperprior entropy model~\cite{minnen2018joint}.

We now give formal description for our probabilistic model. In this section, we assume every variable is a scalar in order to simplify the presentation. In implementation, we apply all the operations element-wise.

\textbf{Posteriors:}
The (approximate) posterior for the $i$-th latent variable, $Z_i$, is defined to be a uniform distribution:
\begin{equation}
\label{eq:method_posterior}
\qzi(\cdot \ | z_{<i},x) \triangleq U(\mu_i - \frac{1}{2}, \mu_i + \frac{1}{2}),
\end{equation}
where $\mu_i$ is the output of the posterior branch in the $i$-th \ourblock{} (see Fig.~\ref{fig:main_train}). From Fig.~\ref{fig:main_overall} and Fig.~\ref{fig:main_train}, we can see that $\mu_i$ (and thus $z_i$) depends on the image $x$ as well as previous latent variables $z_{<i}$.

\textbf{Priors:}
The prior distribution for $Z_i$ is defined as a Gaussian convolved with a uniform distribution~\cite{balle18hyperprior}:
\begin{equation}
\label{eq:method_prior}
\begin{aligned}
\pzi(\cdot \ | z_{<i}) &\triangleq \mathcal{N}(\hat{\mu}_i, \hat{\sigma}_i^2) * U(- \frac{1}{2}, \frac{1}{2})
\\
\LRA \pzi(z_i|z_{<i}) &= \int_{z_i-\frac{1}{2}}^{z_i+\frac{1}{2}} \mathcal{N}(t; \hat{\mu}_i, \hat{\sigma}_i^2) \ dt
\end{aligned}
\end{equation}
where $\mathcal{N}(t; \hat{\mu}_i, \hat{\sigma}_i^2)$ denotes the Gaussian probability density function evaluated at $t$. The mean $\hat{\mu}_i$ and scale $\hat{\sigma}_i$ are predicted by the prior branch (see Fig.~\ref{fig:main_train}) in the $i$-th \ourblock{}.
Note that the prior mean $\hat{\mu}_i$ and scale $\hat{\sigma}_i$ are dependent only on $z_{<i}$, but not on image $x$.
Sampling from $p_i(\cdot \ | z_{<i})$ can be done by:
\begin{equation}
\label{eq:method_prior_sampling}
\begin{aligned}
z_i &\leftarrow \hat{\mu}_i + t \cdot (\hat{\sigma}_i w + u),
\end{aligned}
\end{equation}
where $w \sim \mathcal{N}(0, 1)$ and $u \sim U(-\frac{1}{2}, \frac{1}{2})$ are random noise, and $t \in [0,1]$ is a temperature parameter to control the noise level.
When $t=1$, $z_i$ is an unbiased sample from the prior $p_i(\cdot \ | z_{<i})$, and when $t=0$, $z_i$ is simply the prior mean.

\textbf{Data likelihood:}
For lossy compression, the form of the likelihood distribution $p_{X|Z}(\cdot)$ depends on which distortion metric $d(\cdot)$ is used. In general, we define
\begin{equation}
\label{eq:method_likelihood}
p_{X|Z}(x|z) \propto e^{ -\lambda \cdot d(\hat{x}, x) },
\end{equation}
where $\lambda$ is a scalar hyperparameter that we can manually set, and $\hat{x}$ is the final output of the top-down path network (see Fig.~\ref{fig:main_overall}).
Notice that $\hat{x}$ depends on all latent variables $z \triangleq \{z_1, ..., z_N\}$.
In image compression, $d(\cdot)$ is often chosen to be the mean squared error (MSE), in which case the data likelihood forms a (conditional) Gaussian distribution.

\textbf{Training objective:}
Given an image $x$, our training objective is then to minimize the loss function of VAE (Eq.~\ref{eq:vae_loss}), which can be written as:
\begin{equation}
\label{eq:method_loss}
\begin{aligned}
\mathcal{L} &= D_{\text{KL}}(q_{Z|x} \parallel p_Z) + \EqZx{ \log \frac{1}{p_{X|Z}(x | Z)} },
\\ &= \EqZx{ \sum_{i=1}^N \log \frac{1}{\pzi(z_i | z_{<i})} + \lambda \cdot d(x, \hat{x})} + \text{constant},
\end{aligned}
\end{equation}
where the expectation w.r.t. $Z \sim q_{Z|x}$ is estimated by {drawing a sample at each training step}.
The detailed derivation of Eq.~\ref{eq:method_loss} is given in the Appendix~\ref{sec:appendix_loss}.
We can see that the first term in Eq.~\ref{eq:method_loss} corresponds to a continuous relaxation of the test-time bit rate (up to a constant factor $\log_2 e$), and the second term corresponds to the distortion of reconstruction, where we can tune $\lambda$ to balance the trade-off between rate and distortion.
Note that the model parameters should be optimized independently for each $\lambda$. That is, we use {separately trained models} for different bit rates.

\subsection{Compression}
\label{sec:method_compression}
In actual compression and decompression, the overall framework (\ie, Fig.~\ref{fig:main_overall}) is unchanged, except the latent variables are quantized for entropy coding.
Similar to the Hyperprior model~\cite{balle18hyperprior}, we quantize $\mu_i$ instead of sampling from the posterior, and we discretize the prior to form a discretized Gaussian probability mass function.

The compression process is detailed in Fig.~\ref{fig:main_encode}, in which the \textit{residual rounding} operation is defined as follows:
\begin{equation}
z_i \leftarrow \hat{\mu}_i + \lfloor \mu_i - \hat{\mu}_i \rceil,
\end{equation}
where $\lfloor \cdot \rceil$ is the nearest integer rounding function.
That is, we quantize $\mu_i$ to its nearest neighbour in the set $\{ \hat{\mu}_i + n \mid n \in \mathbb{Z} \}$, denoted by $z_i$, which we can encode into bits using the probability mass function (PMF) $P_i(\cdot)$:
\begin{equation}
P_i(n) \triangleq p_{i}(\hat{\mu}_i + n| Z_{<i}), n \in \mathbb{Z}.
\end{equation}
It can be shown that $P_i(\cdot)$ is a valid PMF, or specifically, a discretized Gaussian~\cite{balle18hyperprior}.
Note that each of our \ourblock{} produces a separate bitstream, so a compressed image consists of $N$ bitstreams, corresponding to the $N$ latent variables $z_1, z_2, ..., z_N$.
In the implementation, we use the range-based Asymmetric Numeral Systems (rANS)~\cite{duda2013ans} for entropy coding.

Decompression (Fig.~\ref{fig:main_decode}) is done in a similar way.
Starting from the constant feature, we iteratively compute $P_i(\cdot)$ for $i = 1,2,...,N$.
At each step, we decode $z_i$ from the $i$-th bitstream using rANS and transform $z_i$ using convolution layers before adding it to the feature.
Once this is done for all $i = 1,2,...,N$, we can obtain the reconstruction $\hat{x}$ using the final upsampling layer in the top-down decoder.

\subsection{Relationship to Previous Methods}
\label{sec:method_discussion}
Our hierarchical architecture (same as ResNet VAEs) may seem similar to the Hyperprior~\cite{balle18hyperprior} model at the first glance, but they are in fact fundamentally different in many aspects.
The inference (encoding) order in the Hyperprior framework is the first order Markov, while our inference is bidirectional~\cite{kingma2016iafvae}.
Similarly, the sampling order of latent variables in the Hyperprior model is also first order Markov,
while in our model, $\hat{X}$ depends on all latent variables $Z$.

A nice property of the ResNet VAE architecture is that, if sufficiently deep, it generalizes autoregressive models~\cite{child2021vdvae}, which are widely adopted in lossy image coders and gives strong compression performance.
Note that this conclusion is not limited to \textit{spatial} autoregressive models.
Other types of autoregressive models, such as channel-wise~\cite{minnen2020channelwise} and checkerboard~\cite{he2021checkerboard} models, can also be viewed as special cases of the ResNet VAE framework.

From the perspective of (non-linear) transform coding~\cite{balle2021nonlinear}, our model transforms image $x$ into $z \triangleq \{z_1, ..., z_N\}$ and losslessly code $z$ instead, so $z$ can be viewed as the transform coefficients (after quantization).
Note that $z$ contains a feature hierarchy at different resolutions, in contrast to many transform coding frameworks where only a single feature is used.
Since $z_i$ is only a quantized version of the posterior mean $\mu_i$, we can also view $\{ \mu_1, ..., \mu_N \}$ as the transform coefficients (before quantization).
Unlike existing methods, there is no separate \textit{entropy model} in our method, and instead, our top-down decoder itself acts as the entropy model for all the transform coefficients $z$.




\section{Experiments}
\label{sec:experiments}


\begin{figure*}[t]
    \centering
    \includegraphics[width=\linewidth]{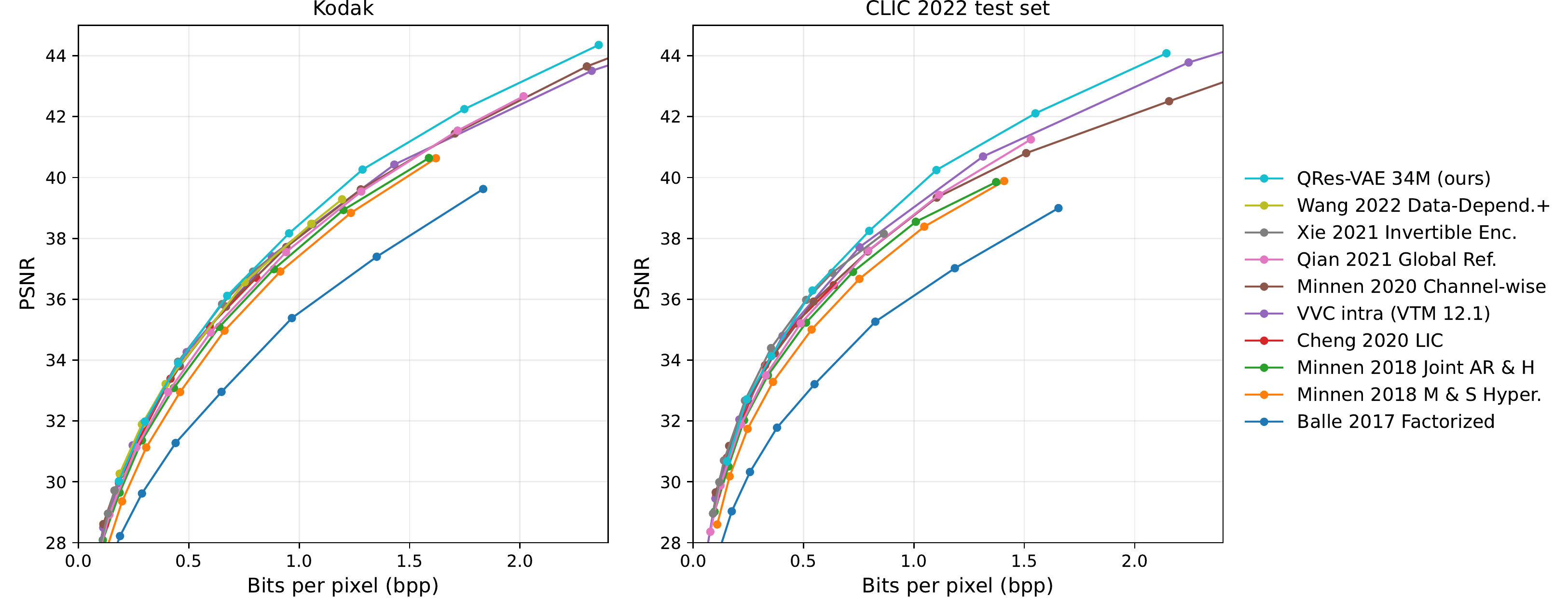}
    \vspace{-0.6cm}
    \caption{\textbf{Lossy compression performance on Kodak (left) and CLIC 2022 test set (right).}
        Our approach is on par with previous methods at low bit rates but outperforms them by a clear margin at higher bit rates.
        BD-rates are shown in Table~\ref{table:exp_complexity}.
    }
    \label{fig:exp_main_lossy}
\end{figure*}
\begin{table*}[ht]
\small
\begin{adjustbox}{width=\linewidth}
\begin{tabular}{l|c|c:c|cc|cc|cc}
\hline
                                                      &                                    &                & Estimated       & \multicolumn{2}{c|}{Latency (CPU)*} & \multicolumn{2}{c|}{Latency (GPU)*} & \multicolumn{2}{c}{BD-rate $\downarrow$} \\
                                                      & Impl.\textsuperscript{\textdagger} & Parameters     & end-to-end FLOPs & Encode           & Decode           & Encode           & Decode           & Kodak               & CLIC\\ \hline
\oursl{} (ours)                                       & -                                  & 34.0M          & \textbf{23.3B}  & 1.124s           & 0.558s           & 0.109s           & 0.076s           & \textbf{-4.076}\%   & \textbf{-4.067}\%  \\ \hdashline
Wang 2022 Data-Depend. +~\cite{wang2022datadependent} & Off.                               & 53.0M          & 523B            & -                & -                & -                & -                & -0.821\%            & -                  \\
Xie 2021 Invertible Enc.~\cite{xie2021invertible}     & Off.                               & 50.0M          & 68.0B           & 2.780s           & 6.343s           & 2.607s           & 5.531s           & -0.475\%            & -2.433\%  \\
Qian 2021 Global Ref.~\cite{qian2021global_ref}       & Off.                               & 35.8M          & 32.1B           & 341s             & 352s             & 35.3s            & 47.6s            & 6.034\%             & 8.133\%            \\
Minnen 2020 Channel-wise~\cite{minnen2020channelwise} & TFC                                & 116M           & 38.1B           & 0.524s           & 0.665s           & 0.136s           & 0.124s           & 0.580\%             & 6.852\%            \\
Cheng 2020 LIC~\cite{cheng2020cvpr}                   & CAI                                & 26.6M          & 60.7B           & 2.231s           & 5.944s           & 2.626s           & 5.564s           & 2.919\%             & 3.300\%            \\
Minnen 2018 Joint AR \& H~\cite{minnen2018joint}      & CAI                                & 25.5M          & 29.5B           & 5.327s           & 9.337s           & 2.640s           & 5.566s           & 10.61\%             & 13.64\%            \\
Minnen 2018 M \& S Hyper.~\cite{balle18hyperprior,minnen2018joint}      & CAI              & 17.6M          & 28.8B           & 0.241s           & 0.442s           & 0.043s           & 0.033s           & 21.03\%             & 27.06\%            \\
Balle 2017 Factorized~\cite{balle2016end2end}         & CAI                                & \textbf{7.03M} & 26.7B           & \textbf{0.204s}  & \textbf{0.411s}  & \textbf{0.040s}  & \textbf{0.032s}  & 67.81\%             & 89.77\%            \\ \hline
\end{tabular}
\end{adjustbox}
{\footnotesize *Latency is the time to code/decode the first Kodak image (W, H=768,512), including entropy coding. CPU is Intel 10700K, and GPU is Nvidia 3090.} \\
{\footnotesize \textsuperscript{\textdagger}We use implementations from authors' official releases (Off.), TensorFlow Compression (TFC)~\cite{tfc_github}, and CompressAI (CAI)~\cite{begaint2020compressai}.}
\caption{\textbf{Computational complexity and BD-rate w.r.t. VVC intra~\cite{pfaff2021vvc_intra} (VTM version 12.1).}
FLOPs are for $256\times 256$ input resolution. Note some values cannot be obtained using author released implementations. Our method outperforms previous methods in terms of BD-rate and at the same time maintains a relatively low computational complexity.
}
\label{table:exp_complexity}
\end{table*}

\subsection{Implementation}
We build two models based on our \ours{} architecture as described in Sec.~\ref{sec:method_architecture}, one with 34M parameters and the other with 17M parameters.
We use the larger model, \oursl{}, for natural image compression, and we use the smaller one \ourss{} for additional experiments and ablation analysis.
Data augmentation, exponential moving averaging, and gradient clipping are applied during training.
We list the full details of architecture configurations in the Appendix~\ref{sec:appendix_detail_architecture} and training hyperparameters in the Appendix~\ref{sec:appendix_training}.

\subsection{Datasets and Metrics}
\textbf{COCO:} We use the COCO~\cite{lin2014coco} \textit{train2017} split, which contains 118,287 images with around $640\times420$ pixels, for training our \oursl{}. We randomly crop the images to $256 \times 256$ patches during training.

\textbf{Kodak:} The Kodak~\cite{kodak} image set is a commonly used test set for evaluating image compression performance. It contains 24 natural images with $768\times 512$ pixels.

\textbf{CLIC:} We also use the CLIC\footnote{http://compression.cc/} 2022 test set for evaluation. It contains 30 high resolution natural images with around $2048\times 1365$ pixels.

\textbf{CelebA (64x64):} We train and test our smaller model, \ourss{}, on the CelebA~\cite{liu2015celeba} dataset for ablation study and additional experiments. CelebA is a human face dataset that contains more than 200k images (182,637 for train/val and 19,962 for test), all of which we have resized and center-cropped to have $64\times 64$ pixels.

\textbf{Metrics:} As a standard practice, we use MSE as the distortion metric $d(\cdot)$ during training, and we report the peak signal-to-noise ratio (PSNR) during evaluation:
\begin{equation}
\text{PSNR} \triangleq -10 \cdot \log_{10} \text{MSE}.
\end{equation}
We measure data rate by bits per pixel (bpp).
To compute the overall metrics (PSNR and bpp) for the entire dataset, we first compute the metric for each image and then average over all images.
We also use the BD-rate metric~\cite{bjontegaard2001bdrate} to compute the average bit rate saving over all PSNRs.

\subsection{Main Results for Lossy Image Compression}
\label{sec:exp_natural}
We compare our method to open-source learned image coders from public implementations.
We use VVC intra~\cite{pfaff2021vvc_intra} (version 12.1), the current best hand crafted codec and a common lossy image coding baseline, as the anchor for computing BD-rate for all methods.
Thus, our reported BD-rates represent average bit rate savings over VVC intra.

Evaluation results are shown in Fig.~\ref{fig:exp_main_lossy}.
We observe that at lower bit rates, our method is on par with previous best methods, while at higher bit rates, our approach outperforms them by a clear margin. We attribute this bias towards high bit rates to the architecture design of our model. We reconstruct the image from a feature map that is $4\times$ down-sampled w.r.t. the original image, which can preserve more high-frequency information of the image.
In contrast, most previous methods aggressively predict the image from a $16\times$ down-sampled feature map, which may be too ``coarse" to preserve fine-grained pixel information.

In Table~\ref{table:exp_complexity}, we compare the BD-rate and computational complexity of our method against previous ones.
For all methods, we use the highest bit rate model for measuring complexity, which represents the worst case scenario.
Because parameter counts and FLOPs are poorly correlated with the actual encoding and decoding latency, we mainly focus on the CPU and GPU execution time for comparison.



Our method outperforms all previous ones in terms of BD-rate on both datasets.
Furthermore, our model runs orders of magnitude faster than the spatial autoregressive models~\cite{minnen2018joint, cheng2020cvpr, qian2021global_ref, qian2021global_ref}, both on CPU and GPU.
Since our approach is fully convolutional and can be easily parallelized, our encoding/decoding latency can be drastically reduced by $10\times$ when a powerful GPU is available, requiring less than $0.1$ seconds to decode a $768\times512$ image.
Note that there are indeed a few baseline methods, such as the Hyperprior~\cite{balle18hyperprior} and Factorized model~\cite{balle2016end2end}, execute faster than our method.
However, this speed difference does not exceed an order of magnitude (\eg, our 0.558s vs. 0.411s~\cite{balle2016end2end} for CPU decoding) and is acceptable given that our method achieves significant better compression efficiency (\eg, our -4.076\% vs. 67.81\%~\cite{balle2016end2end} for the Kodak dataset in terms of BD-rate).



\begin{figure*}[ht]
    \begin{subfigure}[b]{\textwidth}
        \centering
        \includegraphics[width=0.96\linewidth]{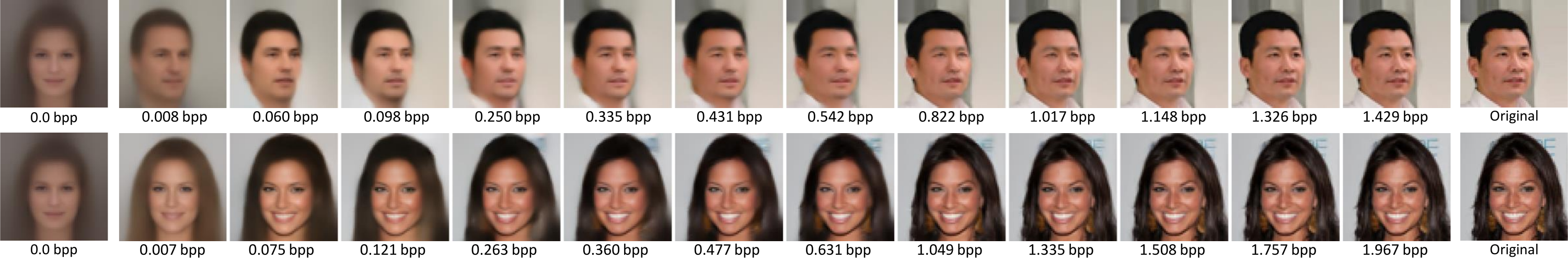}
        \caption{\ourss{}, trained on CelebA, $\lambda=64$. Image resolution is $64\times 64$.}
        \label{fig:exp_progressive_celeb}
        \vspace{+0.2cm}
    \end{subfigure}
    \begin{subfigure}[b]{\textwidth}
        \centering
        \includegraphics[width=0.96\linewidth]{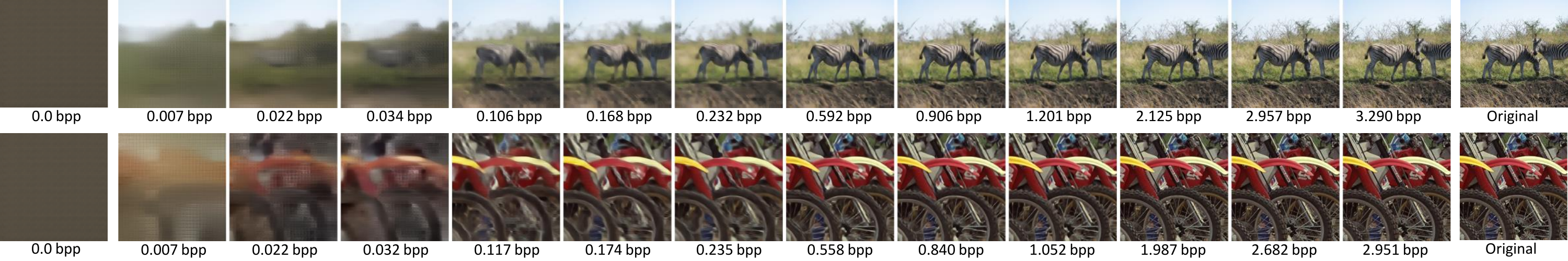}
        \caption{\oursl{}, trained on COCO patches, $\lambda=1024$. Image resolution is $256\times 256$.}
        \label{fig:exp_progressive_natural}
    \end{subfigure}
    \vspace{-0.64cm}
    \caption{\textbf{Examples of progressive decoding}. We decode only the low-dimensional latent variables from the bitstream, and we sample the remaining ones with temperature $t=0$ (\ie, simply using the prior mean). Note that the leftmost images (annotated by 0.0 bpp) can represent the ``average image" in the training dataset.
    Better viewed by zooming in.}
    \label{fig:exp_progressive}
\end{figure*}

Another noticeable difference is that for most previous methods encoding is faster than decoding, while for our \oursl{} this trend is reversed.
Since in most image compression applications, encoding is done only once but decoding is performed many times, our ResNet VAE-based framework shows great potential for real-world deployment.
For example, our model takes 1.124s/0.558s for encoding/decoding on CPU, in contrast to, \eg, the Channel-wise model~\cite{minnen2020channelwise}, which takes 0.524s/0.665s.
This paradigm of higher encoding cost but cheaper decoding of our model is due to the bidirectional inference structure of ResNet VAEs, in which the entire network (both the bottom-up and top-down path) is executed for encoding, but only the top-down path is executed for decoding.


\subsection{Ablative Analysis}
\label{sec:exp_ablation}
To analyze the network design, we train our smaller model, \ourss{}, with different configurations on the CelebA (64x64) training set and compare their resulting losses on the test set.

\textbf{Number of \ourblock{}s.}
We first study the impact of the number of \ourblock{}s, $N$, by varying it from 4 to 12 and compare their testing loss.
For fair comparison, we adjust the latent variable dimensions (\ie, number of channels of $z$) to keep the same total number of latent variable elements. We also adjust the number of feature channels to keep approximately the same computational complexity.
We train models with different $N$ on CelebA (64x64) and compare their testing loss in Table~\ref{table:exp_latent_number}.
All models use $\lambda=32$ (recall that $\lambda$ is the scalar parameter to adjust R-D trade-off).
We observe that with the same number of parameters and FLOPs, deeper model is better, but the improvement is negligible when $N \ge 10$.
We conclude that $N=12$ achieves a good balance between complexity and performance, which we choose as the default setting for our models.

\textbf{Choice of residual network blocks:}
As we mentioned in Sec.~\ref{sec:method_architecture}, the choice of the residual blocks in our network is arbitrary, and different network blocks can be applied in a plug-and-play fashion.
We experiment with three choices: 1) standard convolutions followed by residual connection as in Very Deep VAE~\cite{child2021vdvae}, 2) Swin Transformer block~\cite{liu2021swin}, and 3) ConvNext block~\cite{liu2022convnext}.
Again, for fair comparison, we adjust the number of feature channels to keep a similar computational complexity.
We observe that at $\lambda=32$, the three choices achieve about the same testing loss.
However, when $\lambda=4$, which corresponds to a lower bit rate, the ConvNext block clearly outperforms the alternatives.
We thus use the ConvNext block as our residual block since it achieves a better overall performance.

\begin{table}[t]
\small
\centering
\begin{adjustbox}{width=1\linewidth}
\begin{tabular}{c|ccccc}
\hline
$N$                      & 4      & 6      & 8      & 10     & 12              \\ \hdashline
\# parameters            & 16.7M  & 16.9M  & 16.8M  & 16.9M  & 16.7M           \\
FLOPs (64x64)            & 1.12B  & 1.20B  & 1.22B  & 1.24B  & 1.26B           \\ \hline
Test loss ($\lambda=32$) & 0.2368 & 0.2186 & 0.2121 & 0.2079 & \textbf{0.2078} \\
$\Delta$                & -    & -0.0182 & -0.0065 & -0.0042 & -0.0001        \\ \hline
\end{tabular}
\end{adjustbox}
\caption{Ablation analysis on the number of latent blocks $N$ in \ourss{}, trained on CelebA64.
$\Delta$ is the loss decrement by increasing $N$ by 2.
Deeper model is better, but the amount of improvement decreases as $N$ increases.}
\label{table:exp_latent_number}
\end{table}
\begin{table}[t]
\small
\centering
\begin{adjustbox}{width=1\linewidth}
\begin{tabular}{c|ccc}
\hline
Res. block choice          & Convs~\cite{child2021vdvae} & Swin~\cite{liu2021swin} & ConvNeXt~\cite{liu2022convnext} \\
\hdashline\# parameters    & 17.2M                       & 17.0M                   & 16.7M                           \\
FLOPs (64x64)              & 1.40B                       & 1.38B                   & 1.26B                           \\ \hline
Test loss ($\lambda=4$) \; & 0.0783                      & 0.0790                  & \textbf{0.0757}                 \\
Test loss ($\lambda=32$)   & \textbf{0.2072}             & 0.2077                  & 0.2078                          \\ \hline
\end{tabular}
\end{adjustbox}
\caption{Ablation analysis on the choice of the residual block in \ourss{}, trained on CelebA64
. Overall, ConvNext~\cite{liu2022convnext} blocks give a better performance.}
\label{table:exp_block}
\end{table}

\begin{figure*}[ht]
    \centering
    \begin{subfigure}[b]{0.401\textwidth}
        \centering
        \includegraphics[width=\linewidth]{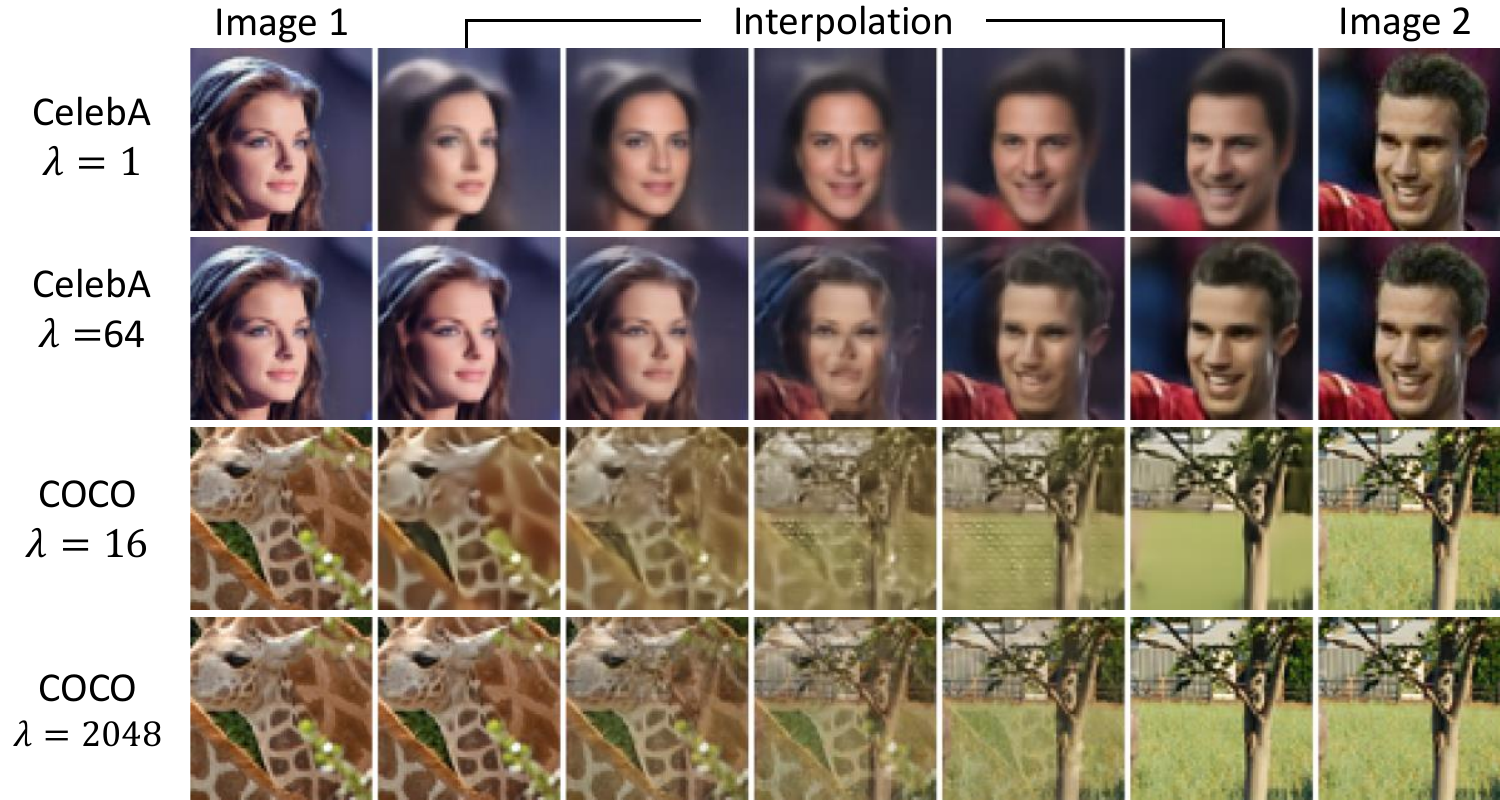}
        \caption{Latent space interpolation}
        \label{fig:exp_qualitative_interpolation}
    \end{subfigure}
    \hfill
    \begin{subfigure}[b]{0.294\textwidth}
        \centering
        \includegraphics[width=\linewidth]{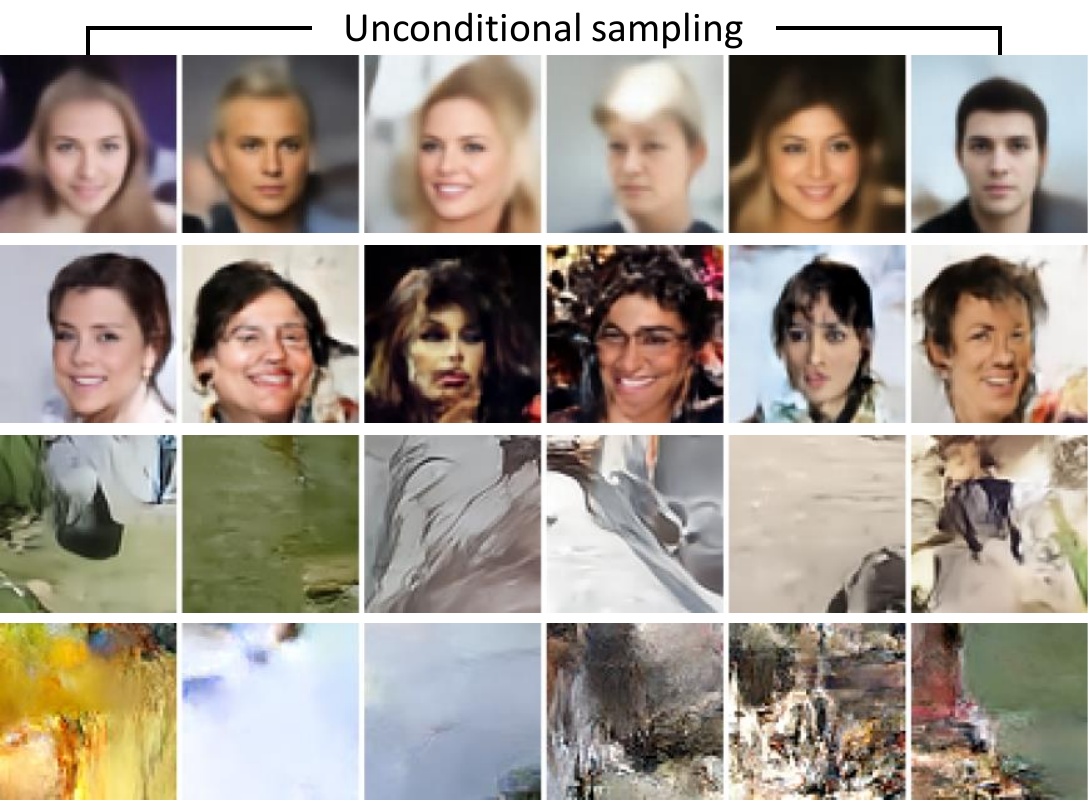}
        \caption{Unconditional sampling}
        \label{fig:exp_qualitative_sampling}
    \end{subfigure}
    \hfill
    \begin{subfigure}[b]{0.294\textwidth}
        \centering
        \includegraphics[width=\linewidth]{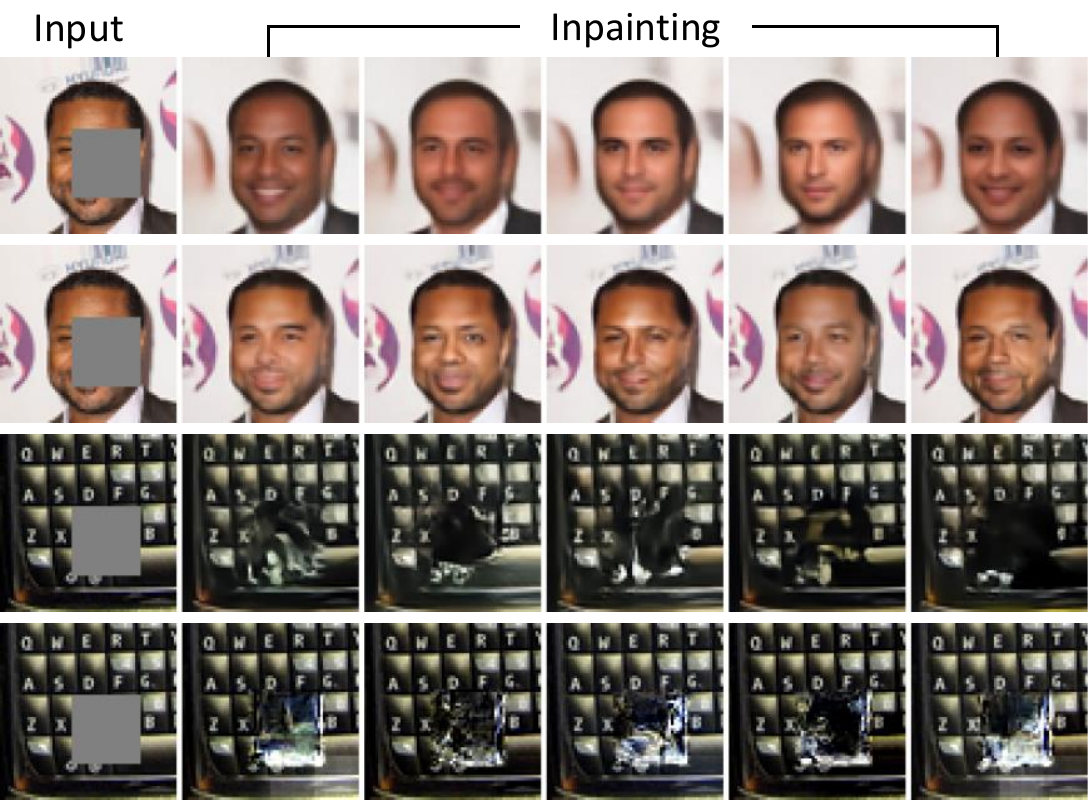}
        \caption{Image inpainting}
        \label{fig:exp_qualitative_inpainting}
    \end{subfigure}
    \caption{\textbf{Analysis of the latent space}.
    Image resolution is $64\times 64$.
    {A lower $\lambda$ represents a model with lower bit rate.}
    We note that our \ours{} learns a semantically meaningful latent space when trained on CelebA with a small $\lambda$ while failed in other settings.
    Despite the weakness when considered as a generative model, our model takes a step towards unifying compression and generative modeling.
    See text for detailed discussion.
    }
    \label{fig:exp_qualitative}
\end{figure*}

\subsection{Additional Experiments}
To further analyze our model, we provide a number of additional experiments in the Appendix, including bit rate distribution (Appendix~\ref{sec:appendix_bpp_distribution}), generalization to different resolutions (Appendix~\ref{sec:appendix_resolution_scaling}), MS-SSIM~\cite{wang2004msssim} training (Appendix~\ref{sec:appendix_msssim}), and lossless compression (Appendix~\ref{sec:appendix_lossless}). We refer interested readers to these sections for more details.

\subsection{Latent Representation Analysis}
\label{sec:exp_generation}
In this section, we show visualization results to analyze how \ours{} represents images in the latent space.

\textbf{Progressive decoding:} We show examples for progressive image decoding in Fig.~\ref{fig:exp_progressive}. 
In Fig.~\ref{fig:exp_progressive_celeb}, we can observe that our method is able to reconstruct the gender, face orientation, and expression of the original face image given only a small subset of bitstreams ($<0.1$ bpp), which correspond to low-dimensional latent variables.
This suggests that the low-dimensional latent variables can encode high-level, semantic information, while the remaining high-dimensional latent variables mainly encode low-level, pixel information.
We also note a similar fact for natural image patches (Fig.~\ref{fig:exp_progressive_natural}): the low-dimensional latent variables encode the global image structure with only a small amount of bits. However, due to the complexity of the natural image distribution, semantic information is much harder to parse. 

\textbf{Latent space interpolation}.
We encode different images into latent variables (without additive uniform noise), linearly interpolate between their latent variables, and decode the latent interpolations.
Results are shown in Fig.~\ref{fig:exp_qualitative_interpolation}.
We observe that when trained on CelebA with $\lambda=1$ (\ie, the lowest bit rate), our model learns a semantically meaningful latent space, in the sense that a linear interpolation in the latent space corresponds to a semantic interpolation in the image space.
However, when $\lambda=64$, latent space interpolation becomes equivalent to pixel space interpolation, indicating that the latent representation mainly carries pixel information.
A similar pixel space interpolation pattern can be observed from the results of our COCO model, suggesting that the COCO model does not learn to extract semantic information, but instead the pixel information.


\textbf{Unconditional sampling}.
Our model can unconditionally generates images in the same way as VAEs, and the samples could give a visualization of how well our model learns the image distribution.
If it learns well, the samples should look similar to real face images (for the CelebA model) or natural image patches (for the COCO model).
The results are shown in Fig.~\ref{fig:exp_qualitative_sampling}.
We observe that the CelebA model generates blurred images when $\lambda=1$ and artifacts when $\lambda=64$, indicating that the model learns global structure but not the pixels arrangements.
Similarly, the COCO model generates samples with some global consistency, but still being different from natural image patches.

\textbf{Image inpainting}.
We also show results on image inpainting in Fig.~\ref{fig:exp_qualitative_inpainting} (the algorithm used is similar to~\cite{lugmayr2022repaint} and is given in our code).
The visualization of inpainting, again, reflects how well our model learns the image distribution.
We obtain a similar conclusion as in previous experiments that our model succeeds in learning the image distribution when trained on the simpler, human face images, while the reconstruction quality degrades for natural image patches.

\section{Conclusion}
In this paper, we propose a new lossy image coder based on hierarchical VAE architectures.
We show that with a quantization-aware posterior and prior model, hierarchical VAEs can achieve state-of-the-art lossy image compression performance with a relatively low computational complexity.
Our work takes a step closer towards semantically meaningful compression and a unified framework for compression and generative modeling.
However, like most learned lossy coders, our method requires a separately trained model for each bit rate, making it less flexible for real-world deployment. 
This could potentially be addressed by leveraging adaptive quantization as used in traditional image codecs, which we leave to our future work.





\clearpage

{\small
\bibliographystyle{ieee_fullname}
\bibliography{arxiv.bib}
}

\clearpage

\section{Appendix}
\label{sec:appendix}

\begin{figure*}[ht]
    \centering
    \begin{subfigure}[b]{\linewidth}
        \centering
        \includegraphics[width=\linewidth]{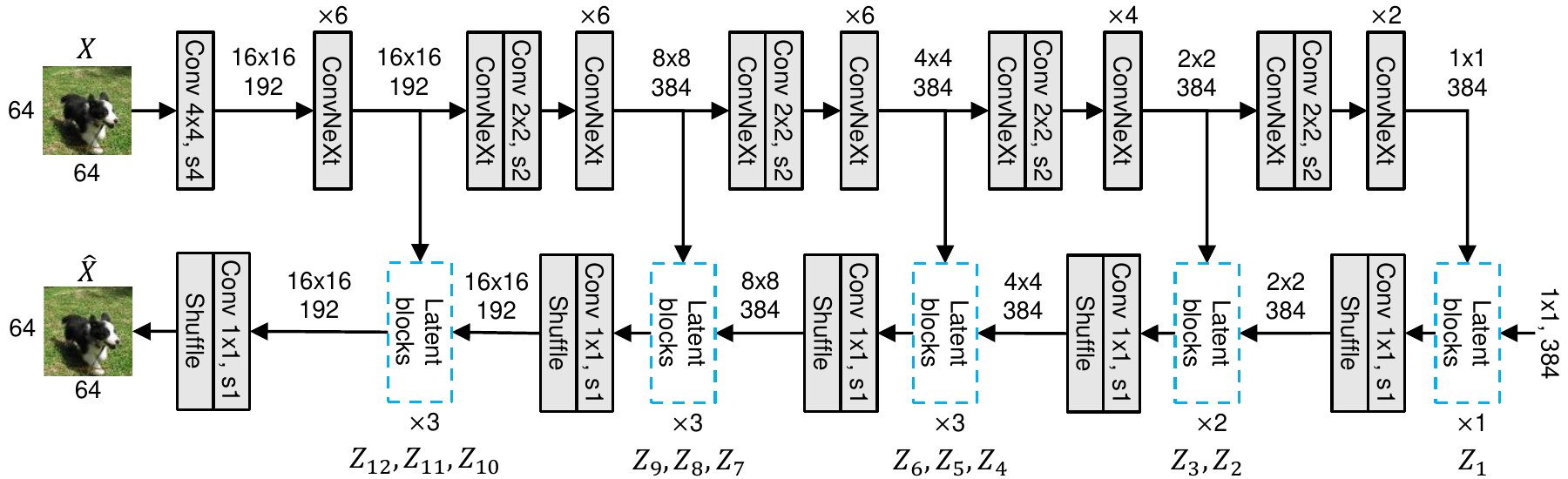}
        \caption{\oursl{}. We train this model on COCO dataset and use it for natural image compression.}
        \label{fig:appendix_architecture_large}
    \end{subfigure}
    \begin{subfigure}[b]{\linewidth}
        \centering
        \includegraphics[width=\linewidth]{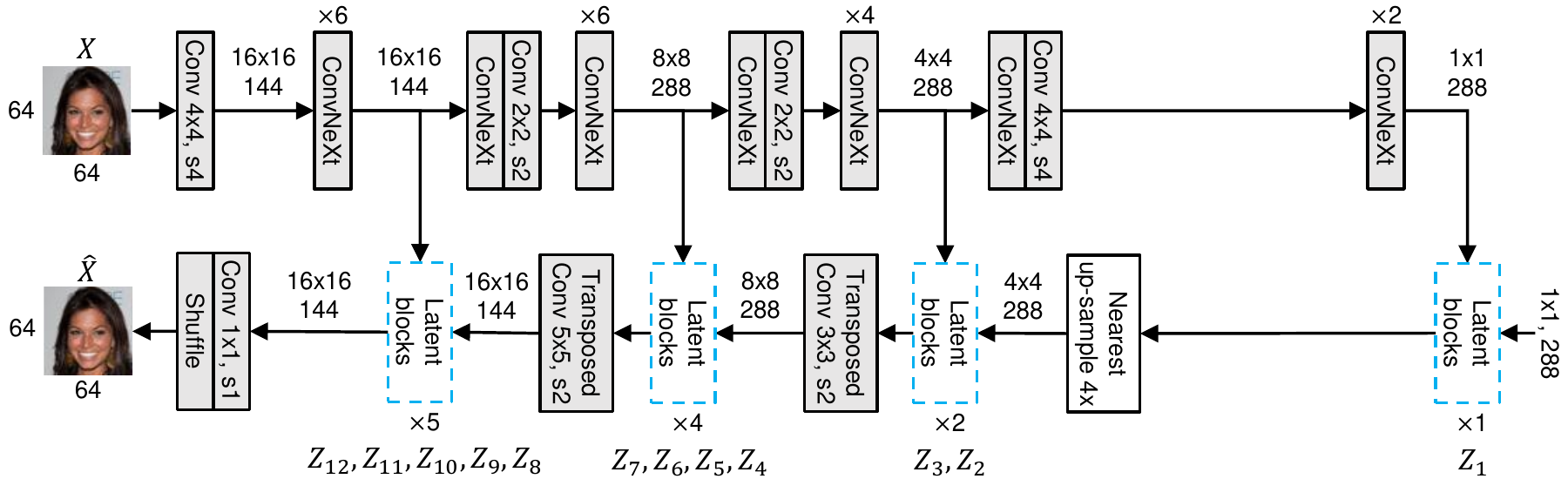}
        \caption{\ourss{}. We train this model on CelebA dataset (human face images) and use it for ablation study.
        Note that in addition to \textit{pixel shuffle}, we also use \textit{nearest upsampling} and \textit{transposed convolutions} for feature map upsampling. The choices of these upsampling operations are arbitrary and do not visibly impact the model performance.
        We adopt this combination in early development and did not further tune them.}
        \label{fig:appendix_architecture_small}
    \end{subfigure}
    \caption{\textbf{Detailed architecture of \ours{}} (64x64 input as an example).
    Numbers on the arrows denote the feature map dimension. For example, ``16x16 192" means a feature map that is $16\times 16$ in height and width, with 192 channels. \textit{``Conv 4x4, s4"} means a convolutional layer with kernel size $4\times 4$ and stride 4. \textit{``ConvNeXt"}~\cite{liu2022convnext} is a modern residual block with depth-wise convolution, layer normalization, linear layers, and GeLU activation. \textit{``Shuffle"} is the pixel shuffle operation~\cite{shi2016subpixel} as often used in super resolution methods.
    }
    \label{fig:appendix_architecture}
\end{figure*}

\subsection{Detailed Architecture}
\label{sec:appendix_detail_architecture}
We show the detailed architecture of \oursl{} in Fig.~\ref{fig:appendix_architecture}, in which we assume the input resolution is $64\times 64$ as an example.
Detailed description of network blocks is provided in the caption.

Since our model is fully convolutional, the input image size is arbitrary as long as both sides are divisible by 64 pixels.
This requirement is because our model downsamples an image by a ratio of 64 at maximum.
In practical cases where the input resolution is not divisible by 64 pixels, we pad the image on the right and bottom borders using the edge values (also known as the \textit{replicate padding}) to make both sides divisible by 64 pixels.
When computing evaluation metrics, we crop the reconstructed image to the original resolution (\ie, the one before padding).

Also note that our model has a learnable constant at the beginning of the top-down path (\ie, on the right most side of Fig.~\ref{fig:appendix_architecture_large} and Fig.~\ref{fig:appendix_architecture_small}).
The constant feature has a shape of $1 \times 1 \times C$, where $C$ denotes the number of channels, for $64 \times 64$ images.
When the input image is larger, say, $256 \times 256$, we simply replicate the constant feature accordingly, \eg, $4 \times 4 \times C$ in this case.

\subsection{Loss Function}
\label{sec:appendix_loss}
Recall that the training objective of VAE is to minimize the variational upper bound on the data log-likelihood:
\begin{equation}
\begin{aligned}
\mathcal{L} &= D_{\text{KL}}(q_{Z|x} \parallel p_Z) + \EqZx{ \log \frac{1}{p_{X|Z}(x | Z)} }
\\
&= \EqZx{ \log \frac{q_{Z|X}(Z|x)}{p_Z(Z)} + \log \frac{1}{p_{X|Z}(x | Z)}  },
\end{aligned}
\end{equation}
where $x$ is a given image in the training set.
In the ResNet VAE, the posterior and prior has the form:
\begin{equation}
\begin{aligned}
q_{Z|X}(z|x) &\triangleq q_{Z|X}(z_1, ..., z_N | x)
\\
&= q_N(z_N|z_{<N},x) \cdots  q_1(z_1|x)
\\
p_{Z}(z) &\triangleq p_{Z}(z_1, ..., z_N)
\\
&= p_N(z_N|z_{<N}) \cdots p_1(z_2|z_1) p_1(z_1),
\end{aligned}
\end{equation}
Plug this into the VAE objective, we have
\begin{equation}
\begin{aligned}
\mathcal{L} &= \EqZx{ \sum_{i=1}^N \log \frac{\qzi(z_i | z_{<i},x)}{\pzi(z_i | z_{<i})} + \log \frac{1}{p_{X|Z}(x | Z)} }.
\end{aligned}
\end{equation}
In our model, the posteriors $\qzi$ are uniform, so on the support of the PDF $q_{Z|x}$, we have $\qzi(z_i | z_{<i},x) = 1, \forall i$.
Also recall our likelihood term $p_{X|Z}(x | Z) \propto e^{ -\lambda \cdot d(\hat{x}, x)}$. Putting them altogether, we have
\begin{equation}
\mathcal{L} = \EqZx{ \sum_{i=1}^N \log \frac{1}{\pzi(z_i | z_{<i})} + \lambda \cdot d(x, \hat{x})} + \text{constant}.
\end{equation}

\subsection{Training Details}
\label{sec:appendix_training}
Detailed hyperparameters are listed in Table.~\ref{table:appendix_hyp_param},
where ``h-clip" denotes random horizontal flipping, and EMA stands for weights exponential moving averaging.
In our experiments, we find that the learning rate and gradient clip should be set carefully to avoid gradient exploding, possibly caused by the large KL divergence (\ie, high bit rates) when the prior fails to match the posterior.

\begin{table}[t]
\small
\centering
\begin{tabular}{l|c|c}
               & \ourss             & \oursl              \\ \hline
Training set   & CelebA train       & COCO 2017 train     \\
\# images      & 182,637            & 118,287             \\
Image size     & 64x64              & Around 640x420      \\ \hdashline
Data augment.  & -                  & Crop, h-flip        \\
Train input size & 64x64              & 256x256           \\ \hdashline
Optimizer      & Adam               & Adam                \\
Learning rate  & $2 \times 10^{-4}$ & $2 \times 10^{-4}$  \\
LR schedule    & Constant           & Constant            \\
Weight decay   & 0.0                & 0.0                 \\ \hdashline
Batch size     & 256                & 64                  \\
\# epochs      & 200                & 400                 \\
\# images seen & 36.5M              & 47.3M               \\ \hdashline
Gradient clip  & 5.0                & 2.0                 \\
EMA            & 0.9999             & 0.9999              \\ \hdashline
GPUs           & 4 $\times$ 1080 ti & 4 $\times$ Quadro 6000 \\
Time           & 20h                & 85h                 \\
\end{tabular}
\caption{Training hyperparameters.}
\label{table:appendix_hyp_param}
\end{table}

\subsection{Rate-Distortion Performance}
\label{sec:appendix_rd}
We provide the numbers of our rate-distortion curves in Table~\ref{table:appendix_kodak_rd} and Table~\ref{table:appendix_clic_rd}, where we show both the estimated, theoretical bit rate (computed from the KL divergence) as well as the actual bit rate (after entropy coding).
We note that on Kodak images (Table~\ref{table:appendix_kodak_rd}), our actual bit rate results in an overhead when compared to the estimated bit rate. This overhead remains approximately a constant of about 0.003 bpp for all bit rates.
In experiments, we find that this overhead is rooted in the entropy coding algorithm, which constantly uses extra 64 bits in each bitstream. Since our model produces 12 bitstreams for each image, for Kodak images ($512 \times 768$) we obtain a constant bpp overhead of
\begin{equation}
    \frac{12 \times 64 \ \text{bits}}{512 \times 768} \approx 0.002 \ \text{bits},
\end{equation}
which approximately matches the overhead we observed in Table~\ref{table:appendix_kodak_rd}.
Note that as the image resolution increases, the percentage of the extra bit rate caused by entropy coding will asymptotically decreases to zero.
As we can observe in Table~\ref{table:appendix_clic_rd}, on CLIC images (around $2048\times 1365$), the actual bit rate is very close to the estimated bit rate, sometimes even smaller than estimated ones, for example when $\lambda = 2048$.

Note that the reported results of \ours{} in the main experiments are the actual bit rates, which is computed after entropy coding.

\begin{table*}[ht]
\centering
\small
\begin{tabular}{l|cccccccc}
$\lambda$            & 16      & 32      & 64      & 128     & 256     & 512     & 1024    & 2048    \\ \hline
Bpp (estimated)      & 0.17960 & 0.29680 & 0.44780 & 0.66960 & 0.94993 & 1.28291 & 1.74430 & 2.35219 \\
Bpp (entropy coding) & 0.18352 & 0.30125 & 0.45200 & 0.67388 & 0.95406 & 1.28697 & 1.74814 & 2.35659 \\
PSNR                 & 30.0210 & 31.9801 & 33.8986 & 36.1126 & 38.1649 & 40.2613 & 42.2478 & 44.3549
\end{tabular}
\vspace{-0.24cm}
\caption{Rate-distortion performance of \oursl{} on Kodak images.}
\label{table:appendix_kodak_rd}
\end{table*}
\begin{table*}[ht]
\centering
\small
\begin{tabular}{l|cccccccc}
$\lambda$            & 16      & 32      & 64      & 128     & 256     & 512     & 1024    & 2048    \\ \hline
Bpp (estimated)      & 0.15370 & 0.24236 & 0.35435 & 0.54013 & 0.79785 & 1.10251 & 1.55179 & 2.14681 \\
Bpp (entropy coding) & 0.15405 & 0.24315 & 0.35457 & 0.54065 & 0.79773 & 1.10183 & 1.55027 & 2.14379 \\
PSNR                 & 30.6719 & 32.7126 & 34.1318 & 36.2879 & 38.2443 & 40.2436 & 42.1072 & 44.0814
\end{tabular}
\vspace{-0.24cm}
\caption{Rate-distortion performance of \oursl{} on CLIC 2022 test set.}
\label{table:appendix_clic_rd}
\end{table*}

\subsection{Comparison with theoretical bounds}
\begin{figure}[ht]
    \centering
    \includegraphics[width=\linewidth]{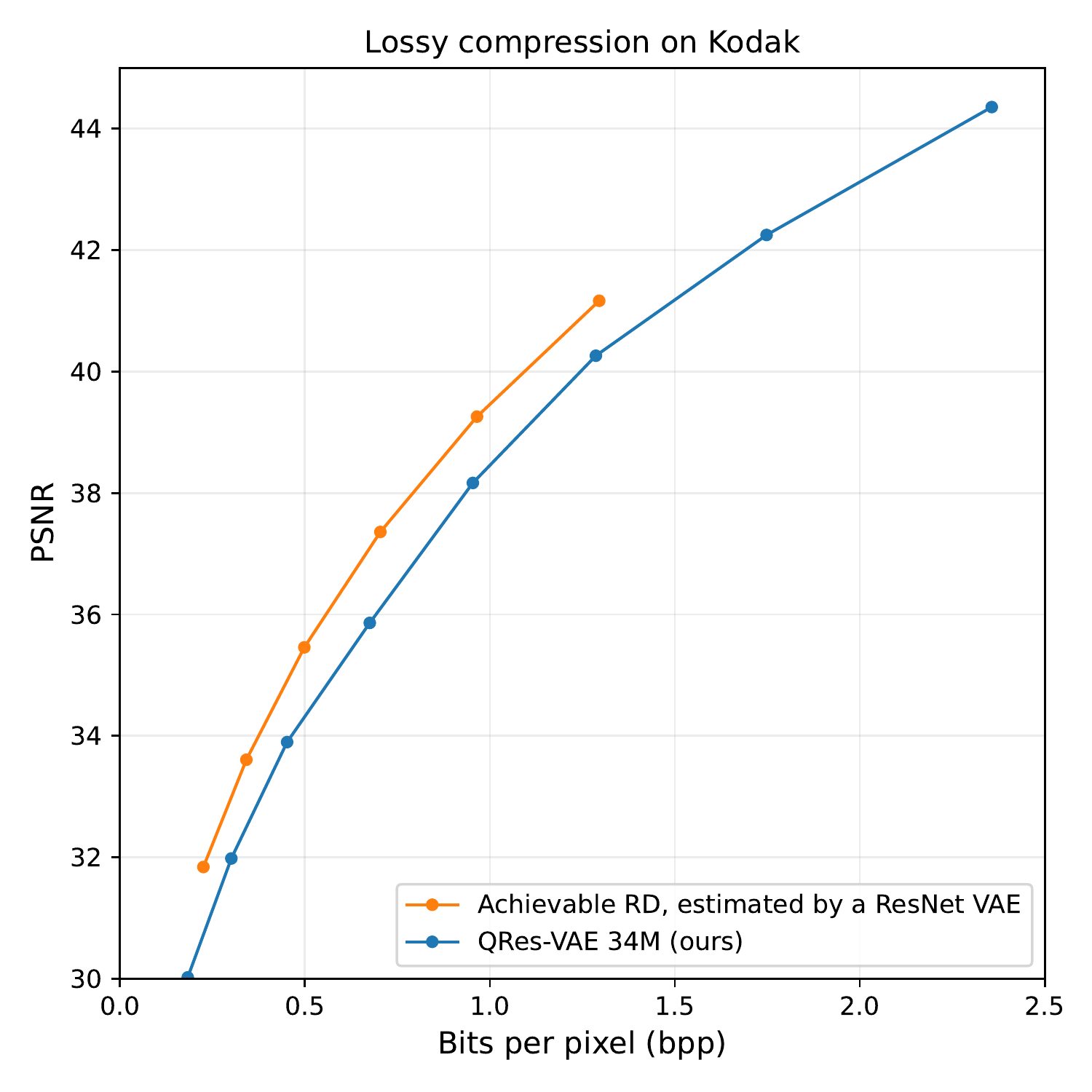}
    \caption{Comparing \oursl{} with the an achievable PSNR-rate curve computed by ResNet VAE (blue line), taken from~\cite{yang2022sandwich}. We can observe that there is still a large room for improvement (around 1dB at all bit rates) for lossy coders.}
    \label{fig:appendix_rd_sandwich}
\end{figure}

As we introduced in the related works, VAEs can be used for computing upper bounds on the information R-D function of images.
We compare our method with one such upper bound, which is computed by a ResNet VAE by Yang \etal~\cite{yang2022sandwich}, in Fig.~\ref{fig:appendix_rd_sandwich}.
The blue curve in the figure is an upper bound of the (information) R-D function of Kodak images, and when viewed in the PSNR-bpp plane, it is a lower bound of the optimal achievable PSNR-bpp curve.
We observe that although our approach improves upon previous method, it is still far from (a lower bound of) the theoretical limit, and further research is required to approach the limit of compression.

\subsection{Bit Rate Distribution}
\label{sec:appendix_bpp_distribution}
\begin{figure*}[ht]
    \centering
    \includegraphics[width=\linewidth]{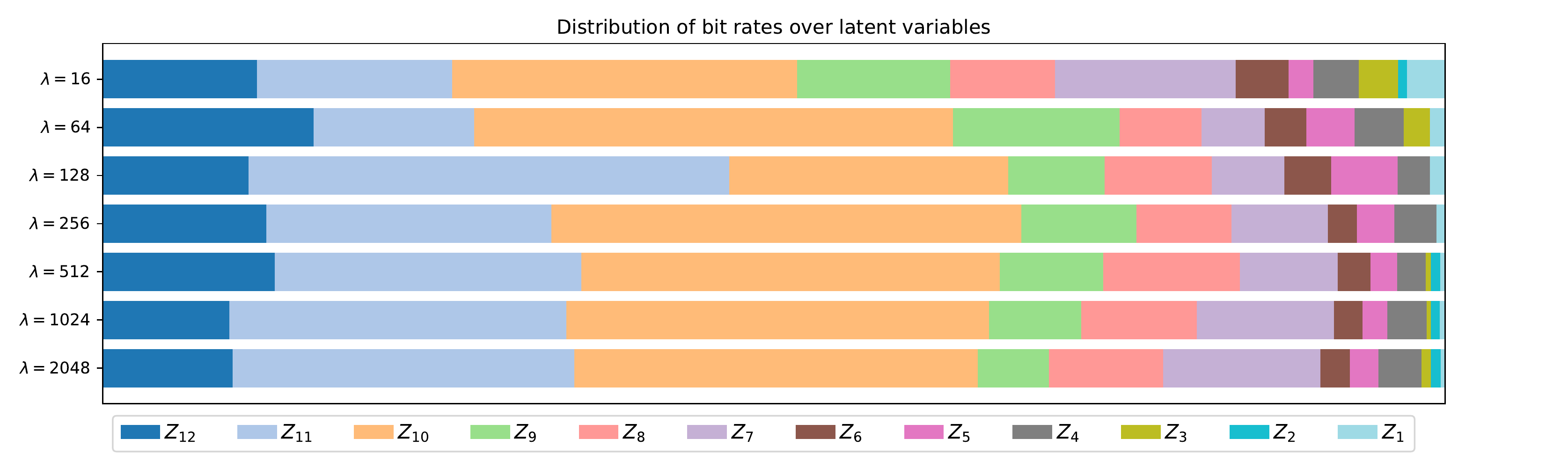}
    \vspace{-0.6cm}
    \caption{\oursl{} bit rate distribution over latent variables.
        Rows correspond to models for different bit rates, \ie, trained with different $\lambda$.
        $Z_1, Z_2, ..., Z_{12}$ are the latent variables, where $Z_1$ has the smallest (spatial) dimension (64$\times$ downsampled w.r.t. the input image) and $Z_{12}$ has the largest dimension (4$\times$ downsampled w.r.t. the input image). See also Fig.~\ref{fig:appendix_architecture_large} for correspondence.
        We observe a similar trend for all models: latent variables with higher dimension cost more bit rates.
    }
    \label{fig:appendix_bpp_distribution}
\end{figure*}

Recall that our model produces 12 bitstreams for each image, where each bitstream corresponds to a separate latent variable.
We visualize how the overall bit rate distributes over latent variables in Fig.~\ref{fig:appendix_bpp_distribution}.
Results are averaged over the Kodak images.
We also notice the posterior collapse, \ie, VAEs learn to ignore latent variables, in our models.
For example, $Z_2$ is ignored by our $\lambda = 64$ model.
Posterior collapse is a commonly known problem is VAEs, and future work need to be conducted to address this issue.

\subsection{Generalization to Various Image Resolutions}
\label{sec:appendix_resolution_scaling}
In our main experiments on natural image compression, we can observe that our model behaves stronger on Kodak than on the CLIC 2022 test set, in the sense that our BD-rate on Kodak is more than 3\% better than the Invertible Enc. model~\cite{xie2021invertible}, while on CLIC this advantage reduces to around 1.5\%.
We hypothesize that this is because our model is trained on COCO dataset, which has a relatively low resolution (around $640 \times 420$ pixels) that better matches Kodak than CLIC.
A better match between training/testing image resolution causes the probabilistic model optimized for the training set also match the testing set, leading to better compression efficiency.

To study this effect quantitatively, we resize the CLIC test set images, whose original resolutions are around $1,365\times 2,048$, such that the longer sides of all images equal to $r$, which we choose from the following resolutions:
\begin{equation}
    r \in \{192, 256, 384, 512, 768, 1024, 1536, 2048\}.
\end{equation}
Then, we evaluate our model as well as two baselines, Minnen 2018 Joint AR \& H~\cite{minnen2018joint} and Cheng 2020 LIC~\cite{cheng2020cvpr}, at each resolution.
Both of the two baselines are obtained from the CompressAI\footnote{github.com/InterDigitalInc/CompressAI} codebase and have been trained on the same high resolution dataset.

\begin{figure*}[ht]
    \centering
    \includegraphics[width=\linewidth]{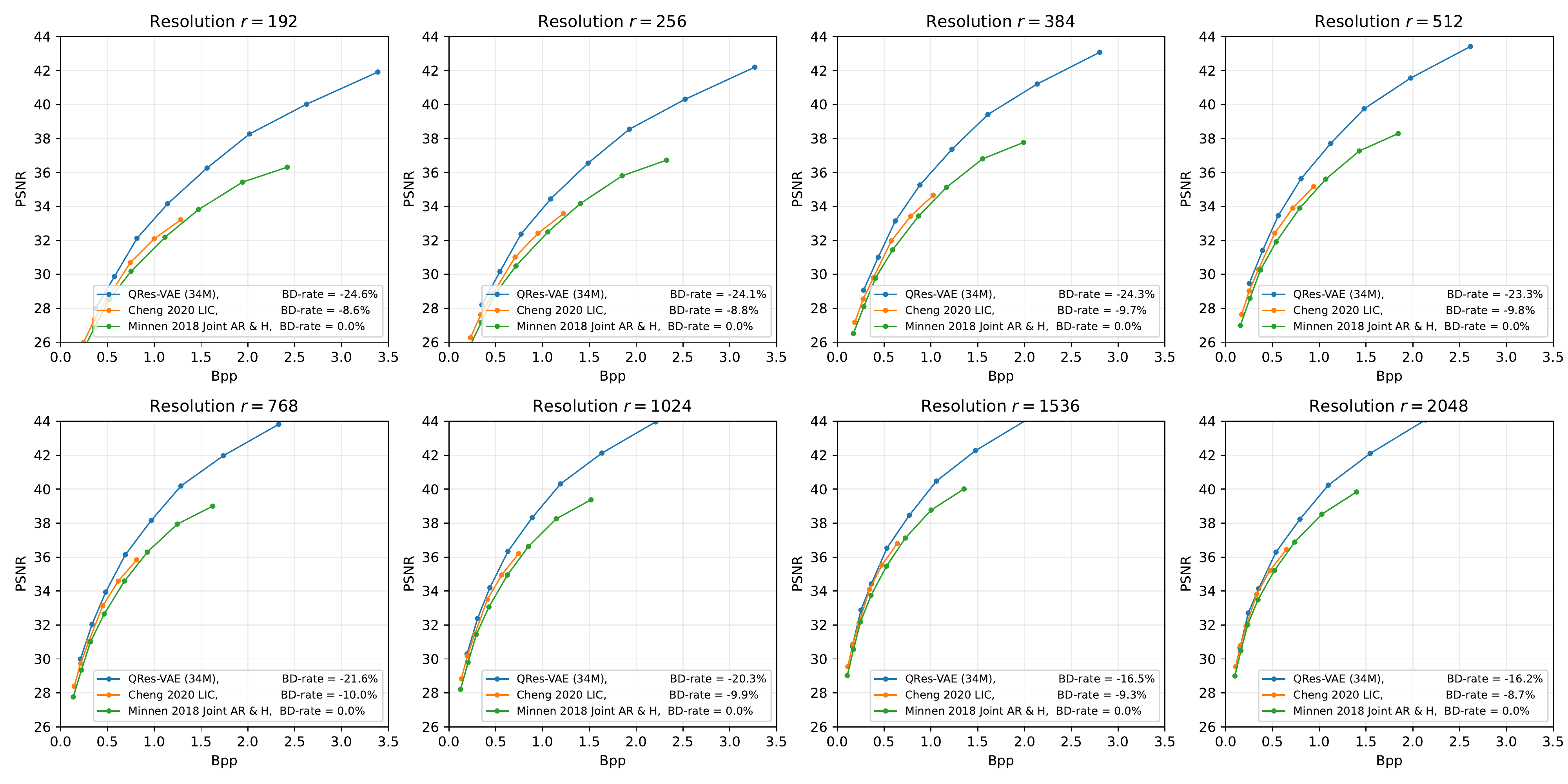}
    \vspace{-0.6cm}
    \caption{We resize the CLIC 2022 test set images such that their longer size equals to $r$ (shown in each subplot), and we evaluate our approach as well as baseline methods on these downsampled images.
    At each resolution, BD-rate is computed w.r.t. the Joint AR \& H model.
    We observe that our \oursl{} is stronger at lower image resolutions.
    }
    \label{fig:appendix_resolutions}
\end{figure*}

Results are shown in Fig.~\ref{fig:appendix_resolutions}.
We observe that as the resolution $r$ increases, the BD-rate of our model w.r.t. the baseline gets worse, from $-24.6\%$ at $r=192$ to $-16.2\%$ at $r=2048$.
In contrast, the BD-rate between two baselines remains relatively unchanged, ranging from $-8.6\%$ to $-10.0\%$.
We thus conclude that \oursl{} model, which is trained on COCO dataset, is stronger at lower resolutions than higher resolutions.
How to design lossy coders that generalize to all resolution images is an interesting but challenging problem, which we leave to future work.

\subsection{MS-SSIM as the distortion metric}
\label{sec:appendix_msssim}
\begin{figure}[t]
    \includegraphics[width=0.98\linewidth]{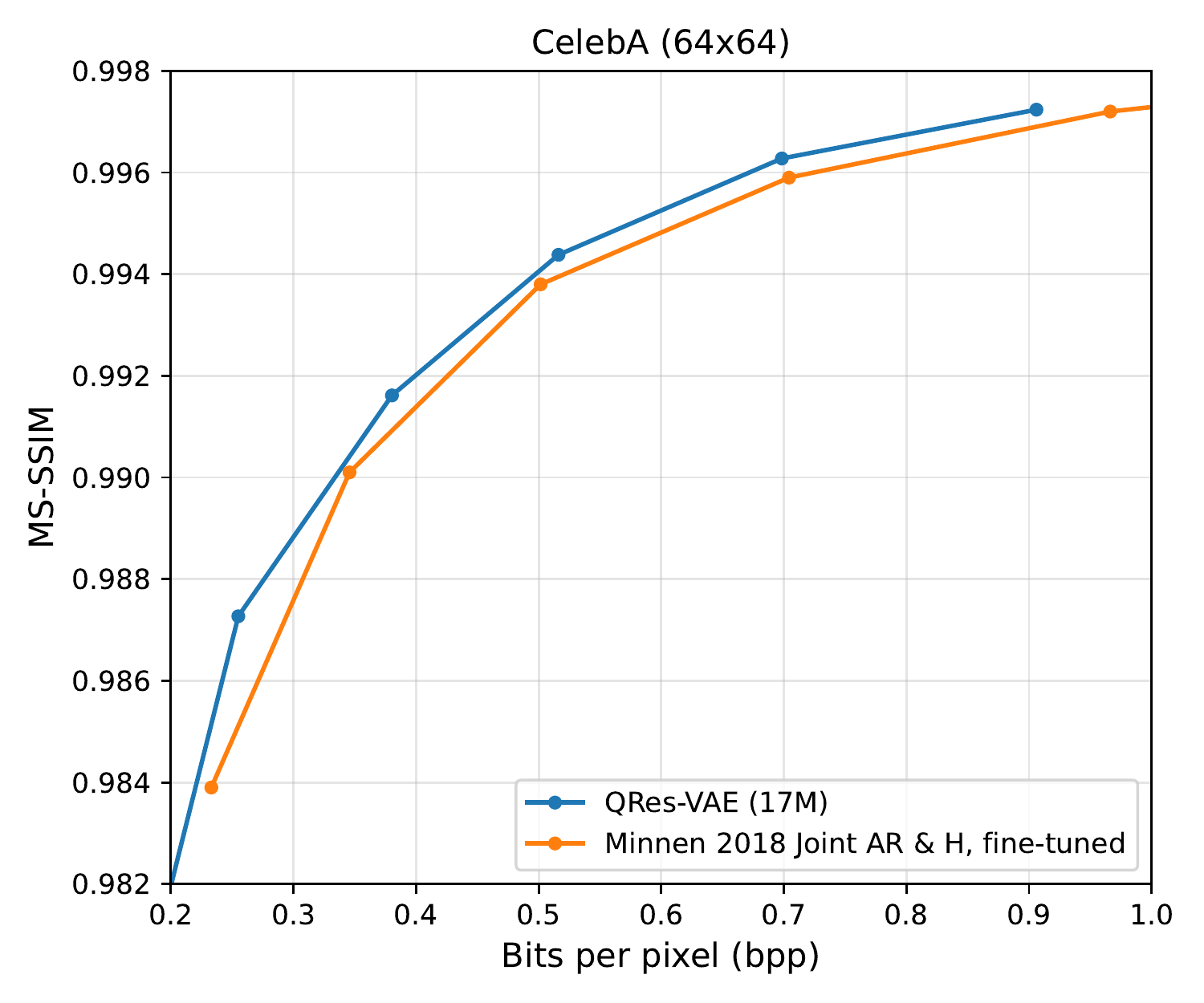}
    \vspace{-0.2cm}
    \caption{MS-SSIM results of \ourss{} on CelebA. Our model is better than the baseline method.
    }
    \label{fig:appendix_celeb64_ms}
\end{figure}

Our model posit no constraints on the distortion metric $d(\cdot)$, so a different metric other than MSE could be used.
For an example, we train our smaller model, \ourss{}, on the CelebA $64\times 64$ dataset to optimize for MS-SSIM~\cite{wang2004msssim}.
Since MS-SSIM is a similarity metric, we let
\begin{equation}
    d(x, \hat{x}) \triangleq 1 - \text{MS-SSIM}(x, \hat{x}).
\end{equation}
Results are shown in Fig.~\ref{fig:appendix_celeb64_ms}, where we also train the Joint AR \& H model~\cite{minnen2018joint} as the baseline.
We observe that our smaller model outperforms the baseline at all bit rates in terms of MS-SSIM, showing that our approach could generalize to using different distortion metrics.

\subsection{Lossless Compression}
\label{sec:appendix_lossless}
Although our method is primarily designed for lossy compression, it can be easily extended to the lossless setting by using a discrete data likelihood term $p_{X|Z}(\cdot)$.
Specifically, we let $p_{X|Z}(\cdot)$ be a discretized Gaussian, which is the same as our prior distributions when at compression/decompression.
Instead of predicting a reconstruction $\hat{x}$ as in lossy compression, we predict the mean and scale (\ie, standard deviation) of the discretized Gaussian using a single convolutional layer at the end of the top-down path, which enables losslessly coding of the original image $x$.

We start from the \oursl{} trained at $\lambda = 2048$, and we fine-tune for lossless compression for 200 epochs on COCO.
Other training settings are the same as \oursl{} shown in Table~\ref{table:appendix_hyp_param}.
We show the lossless compression results in Table~\ref{table:appendix_lossless}.
The \ourslossless{} is better than PNG but is still behind better lossless codecs such as WebP.
We want to emphasize that our network architecture is not optimized for lossless compression, and this preliminary result nevertheless shows the potential possibility of a unified neural network model for both lossy and lossless image compression.

\begin{table}[t]
\small
\centering
\begin{tabular}{l|ccc}
\hline
                & Kodak bpp \\ \hline
PNG             & 13.40     \\
WebP            & 9.579     \\ \hline
\ourslossless{} & 10.37     \\ \hline
\end{tabular}
\caption{Results on lossless compression. Although not designed for lossless compression, our model achieves similar performance compared to common methods.}
\label{table:appendix_lossless}
\end{table}







\end{document}